\documentclass[letterpaper,12pt,leqno]{article}
\usepackage{paper}
\hypersetup{pdftitle={baybutt-2023-crypto_emp_ap_and_high_dim_factor_models}}
\available{http://www.adambaybutt.org/research.html}
\newcommand{\bib}{bibliography.bib}

\newcommand{\mycomment}[1]{}

\begin{document}

\title{Dynamic Latent-Factor Model with \\ High-Dimensional Asset Characteristics}
\author{Adam Baybutt
    \thanks{\textit{Contact}: adam baybutt at protonmail dot com. \textit{Acknowledgments}: I am grateful to Denis Chetverikov for guidance, support, and uniquely valuable mathematical insights. I thank the following people for valuable comments and suggestions: Jesper B\"ojeryd, Mikhail Chernov, Valentin Haddad, Basil Halperin, Jinyong Hahn, Zhipeng Liao, Manu Navjeevan, Andres Santos, Pierre-Olivier Weill, and participants in UCLA's Econometrics Proseminar. Thanks to Pascal Michaillat for formatting. Thank you to Coin Metrics, CoinMarketcap, and Glassnode for providing academic research discounts on data purchases. Replication code available at \url{https://github.com/adambaybutt/crypto_asset_pricing}.
    }}
\date{Last updated: \today}
\begin{titlepage}\maketitle

We develop novel estimation procedures with supporting econometric theory for a dynamic latent-factor model with high-dimensional asset characteristics, that is, the number of characteristics is on the order of the sample size. Utilizing the Double Selection Lasso estimator, our procedure employs regularization to eliminate characteristics with low signal-to-noise ratios yet maintains asymptotically valid inference for asset pricing tests. The crypto asset class is well-suited for applying this model given the limited number of tradable assets and years of data as well as the rich set of available asset characteristics. The empirical results present out-of-sample pricing abilities and risk-adjusted returns for our novel estimator as compared to benchmark methods. We provide an inference procedure for measuring the risk premium of an observable nontradable factor, and employ this to find that the inflation-mimicking portfolio in the crypto asset class has positive risk compensation.

\end{titlepage}

\newpage
\section{Introduction}\label{s:introduction}

In this manuscript, we develop novel estimation procedures with supporting econometric theory for a dynamic latent-factor model with high-dimensional asset characteristics. Additionally, we develop econometric methods for characteristic importance measures in this model and, with asymptotically valid inference, an asset pricing test to measure the risk premium of a nontradable factor. We close with provide empirical results to investigate why different crypto assets earn different average returns; conduct inference for crypto's inflation risk premium; and, estimate risk premia of crypto asset excess returns with classic factor models and our new dynamic latent-factor model.

\paragraph{Factor Model} We assume a statistical model where asset excess returns $r_{i,t+1} \in \R$ are a function of common time-varying latent (unobserved) factors, $f_{t+1} \in \R^k$, as dictated by time-varying asset-specific factor loadings $\b_{i,t} \in \R^k,$ that is, for all assets $i \in \{1, \dots, N\}$ and time $t \in \{1, \dots, T\}$
\begin{equation}\label{e:factor_model}
\begin{aligned}
    r_{i,t+1} &= \a_{i,t} + \b_{i,t}^\top f_{t+1} + \e^r_{i,t+1} \\
    \b_{i,t} &= \G_\b^\top z_{i,t} + \e^\b_{i,t}
\end{aligned}
\end{equation}
where $\a_{i,t} \in \R$ are average pricing errors of the factor model; $\e^r_{i,t+1} \in \R$ are uncorrelated idiosyncratic errors, i.e., $\E_t [ \e^r_{i,t+1} f_{t+1} ] = 0$; $\G_\b \in \R^{p \times k}$ is a static latent loading parameter; and, $z_{i,t} \in \R^p$ are time-varying asset-specific characteristics where $p$ is high-dimensional, i.e., on the order of $N$ and $T.$ Crucially, we follow the established practice in the literature of assuming the number of factors $k$ is low-dimensional, i.e., $N,T,p >> k \in \{1, 2, 3, \dots\}.$ This model has theoretical underpinnings motivated by a structural model for asset excess returns or by the assumption of no arbitrage, as discussed in our literature review.

\paragraph{Double Selection Lasso Factor Model} The main contribution of this research is to develop new estimation and inference procedures, termed the the Double Selection Lasso Factor Model (DSLFM), to fit the latent factors $f_{t+1}$ and loadings $\G_\b$ in \eqref{e:factor_model} and to conduct standard asset pricing tests under the novel setting of high-dimensional asset characteristics.

The DSLFM remains consistent with the equilibrium asset pricing principle that risk premia are solely determined by risk exposures and specifies a linear loading mapping $\G_\b$ between characteristics and dynamic factor loadings $\b_{i,t}.$ We have two novel assumptions for $\G_\b.$ First, we develop estimation procedures and large-sample theory that allows $p,T,N\rightarrow\infty.$ Given our focus is studying the cross-section of crypto assets, this assumption is particularly relevant given the numerous asset characteristics available, as previously discussed, as well as the existence of only a small number of tradable assets and years of relevant data such that $p,T,N$ are of similar order. Second, we assume exact row sparsity in $\G_\b$; that is, only a small number of the $p$ asset characteristics determine the content of the factor loading, which matches empirical findings in cross-sectional asset pricing (\cite{babiak2021risk} and \cite{bianchi2022dynamics}). These novel assumptions within a dynamic latent-factor model require novel estimation procedures and supporting asymptotic theory.

The DSLFM aims to jointly and consistently estimate the loading matrix $\G_\b$ and latent factors $f_{t+1}.$ If we were to utilize the MSE objective function to minimize over the $p-$dimensional choice vector $\G_\b f_{t+1},$ for each $t,$ the mean-squared error of $\sum_{i} (r_{i,t+1} - z_{i,t}^\top \G_\b f_{t+1})^2$, we will have not only a noisily estimated design matrix when $p \sim N,$ (or, at worst, a nonsingular design matrix when $p>N$) but also a non-convex objective function given the interaction between minimization arguments $f_{t+1}$ and $\G_\b$. The next logical step would be to introduce sparsity in $\G_\b,$ which would amount to adding a regularization parameter to the aforementioned objective function to combat the curse of dimensionality from $z_{i,t}.$ However, although potentially helpful for minimizing MSE by decreasing the variance of the estimator, this regularization introduces a bias in estimation, which would lead to invalid asymptotic inference for asset pricing tests, defeating a goal of this research and, more broadly, the purpose of factor models in this field.

We therefore adapt for our purpose the Double Selection Lasso (DSL) estimator developed by \cite{belloni2014inference}. The key insight from their work was to introduce an orthogonality wherein, assuming $\G_\b$ is row sparse, the regularization bias from the LASSO first-stage estimation does not pass through to the target parameter of interest when conducting inference. We first estimate for each time period $t$ and each characteristic $j$ the scalar $\G_{\b,j}^\top f_{t+1}$ using DSL; then, stacking these estimates into a $T \times p$ matrix, we use PCA to obtain separate estimates for latent loadings $\wh{\G}_\b$ and factors $\{ \wh{f}_{t+1} \}_{t=1}^T$; and, finally, we soft-threshold $\wh{\G}_\b$ to set numerous rows to zero given the assumption of sparsity for $\G_\b$. 

This procedure has several additional benefits. Given the period-by-period cross-sectional regression---mirroring Fama-MacBeth regressions--our estimation procedure accommodates unbalanced panels. This first DSL step does require running $T \times p$ cross-sectional regressions; however, this can be done in parallel and is on the order of minutes in practice as each regression is computationally fast. In the second step, the high dimensionality of the PCA procedure, given we have a $p \times T$ matrix, is adapted from the established theory for $N \times T$ excess return matrices \citep{bai2003inferential}. The final soft-thresholding step exploits the sparsity in $\G_\b$ to remove noise from characteristics with low signal-to-noise ratios. \footnote{Finally, just as DSL laid the groundwork for the more general Debiased Machine Learning (DML) theory, this work sets up future research to extend the framework with a semi-parametric specification to utilize the rich set of available machine learning estimators that have been shown to handle well the nonlinearities in cross-sectional asset returns \citep{gu2020empirical}.}

Under standard DSL assumptions adapted to our setting \citep{belloni2014inference}, high-dimensional PCA assumptions \citep{bai2003inferential}, and assuming we observe the true number of latent factors \citep{bai2002determining}, we develop the asymptotic consistency of the latent factors $\wh{f}_{t+1}$ and loading matrix $\Check{\G}_\b$ for the latent factors $f_{t+1}$ and loadings $\G_{\b}$, respectively. Monte Carlo simulations corroborate with finite-sample evidence that the performance of the DSLFM is comparable to or surpasses relevant benchmarks. As is standard in this setting, without further restrictions outlined in \cite{bai2013principal}, $F^0$ and $\G^0_\b$ are not separately identifiable; hence, the $k \times k$ invertible matrix transformation $H$ appears in each asymptotic result. However, in many cases, knowing $F^0 H$ is equivalent to knowing $F^0;$ for example, using the regressor $F^0$ will give the same predictions as using the regressor $F^0 H$ given they have the same column space. Similarly, the target parameter in the coming inference result is rotation invariant to $H.$ 

To show the generality of these estimation procedures, we enrich our model---with one of several possible extensions---to address a common question in asset pricing research. We ask whether an observable, nontradable factor $g_{t+1} \in \R$ carries a risk premium: compensation for exposure to the risk factor holding constant exposure to all other sources of risk, i.e., variation with other factors. In the subsequent empirical applications, we investigate a common hypothesis for the crypto asset class: exposure to inflation offers crypto investors a positive risk premium.

Following a recent approach in the literature (\cite{giglio2021asset} and \cite{giglio2021test}), we assume the ``true'' latent factors $f_{t+1}$ can be decomposed into the latent-factor risk premia $\g \in \R^k$ and latent-factor innovations $v_{t+1} \in \R^k,$ that is, $f_{t+1} \coloneqq \g + v_{t+1}.$ Then, we specify the observable factor $g_{t+1}$ as potentially linearly correlated with the latent factors through
\begin{equation*}
g_{t+1} = \h v_{t+1} + \e^g_{t+1},
\end{equation*}
where $\h \in \R^{k}$ is an unknown parameter mapping the relation between the latent-factor innovations and the observable factors, and $\e^g_{t+1} \in \R$ is measurement error in $g_{t+1}.$

The risk premium of an observable factor---our target parameter in this extension---is defined to be the expected excess return of a portfolio with loading (i.e., beta) of 1 with respect to this factor in $g_{t+1}$ and zero loadings on all other factors; in this model, that parameter is $\g_g \coloneqq \h^\top \g = \h^{0 \top} H H^{-1} \g^0 = \h^{0 \top} \g^0,$ which utilizes the rotation invariant result of \cite{giglio2021asset}.

We thus extend with our estimation procedure for $\g_g$ in a dynamic latent factor model with high-dimensional characteristics the estimation procedure of \cite{giglio2021asset}, which is for a static latent factor model. We additionally develop our estimator's large-sample distribution and variance to conduct asset pricing tests on the sign of the observable factor risk premium. We apply this test for the risk premium of the observable inflation factor within the crypto asset panel studied herein.

\paragraph{Empirical Applications} We close with studying three empirical questions: the out-of-sample pricing ability of the DSLFM compared to benchmark methods, the drivers of returns using our bootstrapped inference method, and estimating the inflation risk premium in the crypto asset class with our asset pricing test. We utilize the same novel and rigorous panel of crypto asset returns developed in XYZ.

To begin, we compare the out-of-sample predictive power in the Q3-Q4 2022 data of three models, namely, a three-factor model of size, crypto market, and momentum; a latent three-factor model fit with PCA; and a dynamic latent-factor model fit with IPCA using a subset of the characteristics. We find IPCA outperforms the other models, suggestive of the signal in the characteristics. Its predictive pricing signal outperforms a random walk and it provides economically and statistically significant risk-adjusted returns for the zero-investment portfolio, whereas the other models underperform a random walk yet provide modest long-short risk-adjusted returns.

Utilizing the broader set of asset characteristics, we first establish the comparable out-of-sample predictive ability of the DSLFM compared to the benchmark methods, with supporting bootstrapped characteristic importance measures to elucidate the drivers of returns. Exchanges inflows and outflows were significant characteristics, showing the importance of these onchain measures. While DSLFM achieved a maximum, with one latent factor, out-of-sample Sharpe of $3.3,$ this underperforms ICPA's maximum, with one latent factor, Sharpe ratio of $4.07$. 

Additionally, we implement our testing procedure to find that the crypto asset class provides investors a positive inflation risk premium. Early proponents of Bitcoin and other cryptocurrencies framed these as an outside options or hedges against traditional fiat currencies. To study this question, we use our extended model to recover the 10-year expected inflation mimicking portfolio and measure its risk premium. This inflation risk premium was estimated at a statistically significant 1.4 basis points (0.0097 standard error). This translates to a 7.3\% annual excess return, suggestive of positive compensation for investors holding an inflation-hedged crypto portfolio, ceteris paribus.

\section{Literature Review}\label{s:litreview}

This paper builds the econometric theory literature for high-dimensional factor models. \cite{giglio2022factor} provide an excellent review of recent machine-learning based factor model applications and relevant econometric theory, including the common asymptotic frameworks of fixed $N$ and $T\rightarrow\infty,$ fixed $T$ and $N\rightarrow \infty,$ and $T,N\rightarrow \infty.$ In short, this paper extends this last asymptotic framework to be high-dimensional on the new dimension of the number of asset characteristics, i.e., $p,T,N \rightarrow \infty$.

Starting from either a structural model for asset excess returns in the style of the Capital Asset Pricing Model \citep{sharpe1964capital}, or the assumption of no arbitrage, as in Arbitrage Pricing Theory \citep{ross1976arbitrage}, a stochastic discount factor $m_{t+1} \in \R$ exists and an Euler equation, termed the Law of One Price, holds for asset excess returns $r_{i,t+1} \in \R$ for assets $i \in \{1,2,\dots,N\}$ in time periods $t \in \{1,2,\dots,T\}$
\begin{equation*}\label{e:loop}
    \E_t [ m_{t+1} r_{i,t+1}] = 0,
\end{equation*}
which by the definition of the variance and covariance operators,
\begin{equation*}\label{e:factor_model_overarching}
    \E_t [ m_{t+1} r_{i,t+1}] = \underbrace{\frac{Cov_t ( m_{t+1}, r_{i,t+1} )}{Var_t \left(m_{t+1} \right)}}_{\b_{i,t}} \underbrace{\frac{- Var_t \left(m_{t+1} \right)}{E_t [m_{t+1}]}}_{\lambda_t}.
\end{equation*}

As discussed in Section \ref{s:introduction} we directly assume the statistical model
\begin{equation*}
    r_{i,t+1} = \a_{i,t} + \b_{i,t}^\top f_{t+1} + \e^r_{i,t+1}.
\end{equation*}
To map this model to its theoretical underpinnings in the Law of One Price, one can assume for all $i$ and $t$: mean zero unobserved idiosyncratic errors $\E_t [\e^r_{i,t+1}] = 0,$ uncorrelated errors $\E_t [ \e^r_{i,t+1} f_{t+1} ] = 0,$ the price of risk associated with the factors to be defined as $\l_t \coloneqq \E_t [ f_{t+1} ],$ and zero average pricing errors $\a_{i,t} = 0$ \citep{cochrane2009asset}. The factor model posits that asset excess returns $r_{i,t+1}$ signify compensation for asset-specific, time-varying exposure $\b_{i,t} \in \R^k$ to systematic risk factors $f_{t+1} \in \R^k.$ 

The classic factor model is a static loading observable factor model---in the style of \citeauthor{fama1992cross}---where pricing errors and static factor loadings $\b_i$ in \eqref{e:factor_model} using exogenously-defined factors are estimated via the ``Fama-MacBeth'' two-step procedure \citep{fama1973risk}, which has rich supporting inferential theory \citep{shanken1992estimation}. This procedure relies on ex ante declaration of observable factors $f_{t+1}$ formed as a convex combination (i.e., a portfolio) of sorted asset returns $r_{t+1}$ based on asset characteristics $z_{t} \in \R^p.$ This is likely an incomplete model of the relationship between $z_{t}$ and $r_{t+1}$ and prone to overfit.

Recognizing the proliferation of factors, a ``Factor Zoo'' \citep{cochrane2011presidential}, explaining the cross section of expected returns, \cite{feng2020taming} propose the use of Double Selection Lasso \citep{belloni2014inference} combined with two-pass Fama-MacBeth regressions to evaluate the contribution of a new factor, $g_{t+1},$ explaining asset expected returns above and beyond an existing high-dimensional set of factors. However, as the recent empirical literature has shown (e.g., \cite{kelly2019characteristics}; \cite{chen2020deep}), allowing the data to construct the relevant latent factors offers superior explanatory and predictive power as compared to using a set of selected observable factors from the literature.

Factor models with latent factors have been a focal point since the development of APT \citep{ross1976arbitrage} and early empirical efforts \citep{chamberlain1983funds}. PCA is the common estimation framework. \cite{bai2003inferential} develops the core theory for PCA estimation and inference under joint $N,T\rightarrow\infty$ high-dimensional asymptotics, with \cite{bai2002determining} introducing a novel Information Criterion penalization to ascertain the true number of latent factors. Assumptions for the joint identification of factor loadings and factors are summarized in \cite{bai2013principal}. This manuscript leverages these contributions yet employs instead a dynamic latent-factor model with $p,N,T\rightarrow\infty$ high-dimensional asymptotics. In the asset pricing context, the static model can still perform well for describing portfolios over time given the dynamics are captured by the dynamic factors; however, this has not performed as well in describing individual asset returns, as noted by \cite{ang2009using}. Although this model allows the data to statistically inform the factor structure, it fails to incorporate rich asset characteristic data and it assumes a static factor loading that maps systematic risks to excess returns. 

A recent and significant methodological advance, instrumented PCA (IPCA) from \cite{kelly2019characteristics}, utilizes asset characteristics in a linear model of dynamic factor loadings to parameterize the more general semi-parametric method studied in \cite{connor2007semiparametric}. That is, 
\begin{equation*}
    \b_{i,t} = \G_\b^\top z_{i,t} + \e^\b_{i,t}
\end{equation*}
where $\G_\b \in \R^{p \times k} $ is a loading matrix mapping asset characteristics to the factor loading. IPCA has several benefits, including compressing the $N \times T$ factor loading matrix $\b$ to a lower dimensional $p \times k$ loading matrix $\G_\b$, as well as specifying a time-varying relationship $\b_{i,t}$ between characteristics and returns, which as stated previously appears to be the empirical reality in crypto cross-sectional asset pricing. In a separate theory paper, the authors develop the asymptotic distributional theory---following $N,T\rightarrow \infty$ asymptotics from \cite{bai2003inferential}---for the factor realizations and loadings under quite general identifying restrictions on loadings and factors \citep{kelly2020instrumented}.

The IPCA procedure benefits from the following: the efficiency gains from using asset characteristics for estimating the latent factors and their loadings; accommodation of unbalanced panels; maintaining an expected return factor model structure to ascertain the economic relationships among factors and assets via the observable characteristics; and, a parametric model with inference procedures for asset pricing tests. The IPCA estimation procedure, however, is not possible under high-dimensional asset characteristics (i.e., $p > T,N$), or, if regularization is used, produces biased inference for asset pricing tests.

\cite{giglio2021asset} develop a three-step procedure combining estimation of latent-factor model via PCA with standard two-pass regressions to recover the risk premium of an observable nontradable factor $g_{t+1} \R$, which are potentially correlated with the latent factors:
\begin{equation*}
    g_{t+1} = \h^\top v_{t+1} + \e^g_{t+1}
\end{equation*}
where $v_{t+1}$ are (mean-zero) latent-factor innovations (i.e., $f_{t+1} = \g + v_{t+1}$); $\h \in \R^{k}$ is a linear mapping of the true latent factors to the observed factors; and, $\e^g_{t+1}$ is measurement error. This allows the observable factors to either be some component of the latent factors or just correlated with $v_{t+1}$ and therefore still carry a risk premium. The target parameter is the risk premium associated with the observable factors $\g_g \coloneqq \h \g.$

To demonstrate the generality of the DSLFM, we extend our estimation procedure by adding the procedure of \cite{giglio2021asset} to address this classic asset pricing test of the recovering the risk premium of an observable factor. The DSLFM theory presented in Section \ref{s:dslfm} extends \cite{giglio2021asset} by incorporating not only dynamic factor loadings, but also high-dimensional asset characteristics.

More recent literature has incorporated a wide array of machine learning-based estimation approaches within the factor model structure. \cite{gu2020empirical} study a set of machine learning estimation procedures for measuring the equity risk premium to find that deep learning methods outperform out-of-sample. \cite{gu2021autoencoder} develop a factor model in the structure of IPCA, but allows for non-linear mappings to the factor loadings and the factors through two feed-forward neural networks. Although, in practice they only use linear mappings to the factors, they still obtain out-of-sample predictive $R^2$ and Sharpe ratio gains in relation to benchmark methods. Other notable uses of deep learning within factor model structures are \cite{chen2020deep}; \cite{feng2018deep}; \cite{guijarro2021deep}; among others.

A fundamental difference between common empirical settings for machine learning applications and their use in empirical asset pricing is the uniquely low signal-to-noise DGP. Theory suggests market efficiency prices in the signal, such that, the unforecastable idiosyncratic error dominates. This critical issue significantly compounds the curse-of-dimensionality of using high-dimensional asset characteristics, which further motivates the parsimonious specification of the factor model.

\section{Double Selection Lasso Factor Model}\label{s:dslfm}

Our goal is to consistently estimate a dynamic latent-factor model and conduct asset pricing tests with valid inference under the novel setting of high-dimensional asset characteristics, which is of particular relevance in the crypto asset class.

\subsection{Setup}\label{s:setup}

\paragraph{Setting and Observable Random Variables} Assume for time periods $t=1,2,\dots,T$ and assets $i=1,2,\dots,N,$ that we observe realizations of random variables for asset excess returns $r_{i,t+1}\in \R$ and asset characteristics $z_{i,t} \in \R^{p}.$ An asset's excess return is the simple return of asset $i$ from time $t$ to $t+1$ net the assumed simple return of the risk-free rate (e.g., one-month US Treasury Bill). An asset characteristic of asset $i$ is known at time $t:$ for example, the total fees for a crypto protocol between time $t-1$ and $t.$ Note that asset characteristics are information from the previous period to follow the established convention in the literature and to be able to use this model for prediction. Importantly, we will introduce the novel asymptotic assumption for dynamic latent-factor models to let $p$ grow to infinity simultaneously with $N$ and $T.$

\paragraph{Model} Given the highly nonlinear data-generating process observed in empirical asset returns (\cite{gu2020empirical}; \cite{chen2020deep}; \cite{bianchi2021bond}), we specify a semi-parametric factor model---where $\b_{i,t}$ is a function of asset characteristics $z_{i,t}$---studied in recent literature (\cite{connor2007semiparametric}; \cite{connor2012efficient}; and \cite{fan2016projected}). We assume a dynamic latent-factor model where
\begin{equation*}
r_{i,t+1} = \b_{i,t}^\top f_{t+1} + \e^r_{i,t+1}.
\end{equation*}

However, given we are interested in conducting inference, we specify a linear model---to build on the foundational work of \cite{kelly2019characteristics}---for the factor loadings 
\begin{equation*}
\b_{i,t} = \G_\b^\top z_{i,t} + \e^\b_{i,t}.\footnote{Although we are working with this parametric specification for the factor loadings, we can in our setting, nevertheless, employ feature engineering to generate many different functional forms of our asset characteristics, given the coming dimensionality reduction from LASSO.}
\end{equation*}

\paragraph{Parameters and Unobserved Random Variables} $f_{t+1} \in \R^k$ are low-dimensional latent factors; $\beta_{i,t} \in \R^k$ are latent-factor loadings; $\G_\b \in \R^{p \times k}$ is an unknown factor loading parameter matrix; and, $\e^r_{i,t+1} \in \R$ and $\e^\b_{i,t} \in \R^k$ are unobserved idiosyncratic errors.

The latent factors $f_{t+1}$ should be interpreted as purely statistical in nature. That is, these risk factors do not necessarily capture fundamental shocks to productive technologies as modeled in canonical theoretical models. Nevertheless, the latent factors capture systemic risk or covariance among asset returns that is non-diversifiable. We follow the literature in restricting $k$ to be a small finite constant (i.e., $k \in \{1,2,3,4,5\}$) that, in our asymptotic theory, does not grow with $p,T,N.$ It should be noted that, although ubiquitous, it is nevertheless a strong assumption: the empirical content of asset returns can be captured by a small number of strictly time-varying systematic risk factors.

The specification of $\b_{i,t}$ provides several benefits. First, we enable the use of a dynamic loading to model a changing relationship (e.g., regime changes) between the cross section of returns and systematic risk. Yet, we reduce the parameter space from a $N \times T$ loading matrix $\beta$ to the $p \times k$ loading matrix $\G_\b.$ Second, we incorporate additional data from the large number of asset characteristics to influence the factor model for returns through the loading matrix $\G_\b$. This addresses a challenge of migrating assets wherein an asset-specific but static $\b_i$ would not be able to capture an asset moving from, for example, a crypto asset earning low fees to one with high protocol fees. The classic way to handle this issue was to sort assets into portfolios of similar characteristics to form test assets, which then compresses the dimensionality of the cross-section. Thus, as discussed in \cite{kelly2019characteristics}, this model specification skips ad hoc test asset formation to instead accommodate working directly with the high-dimensional system of individual assets. Finally, we assume exact row sparsity in $\G_\b$---precisely stated in the coming Assumption \ref{assump_dsl_consist_main}(ii)---a novel assumption to the literature; that is, only a small number of the $p$ asset characteristics determine the content of the factor loading, which matches empirical findings in cross-sectional asset pricing (\cite{kelly2019characteristics}; \cite{babiak2021risk}; \cite{bianchi2022dynamics}).

\paragraph{Extended Model} In order to show the generality of this approach, we enrich the model---with one of several possible extensions---to address the common question in asset pricing research of whether an observable factor $g_{t+1} \in \R$ carries a risk premium: compensation for exposure to the risk factor holding constant exposure to all other sources of risk, i.e., variation with other factors.

In the context of asset pricing, a factor can be either tradable or nontradable. A tradable factor is a portfolio, that is, a convex combination of tradable asset returns. The risk premium is straightforward to calculate for tradable factors: it is the time series average excess return of the factor. However, many risk factors are nontradable, e.g., inflation expectations, consumption, liquidity, etc. Thus, we must estimate the risk premia of nontradable observable risk factors as the risk premia associated with their tradable portfolio.

Following a recent approach in the literature (\cite{giglio2021asset} and \cite{giglio2021test}), first, we assume the aforementioned model for returns is a function of the ``true'' latent factors $f_{t+1},$ \footnote{To be precise, by true latent factors, we mean we can consistently estimate the finite constant of the number of latent factors that span the cross section of returns.} and, second, we assume these true latent factors can be decomposed into the latent-factor risk premia $\g \in \R^k$ (i.e., unknown parameters of the long-run average excess return) and latent-factor innovations $v_{t+1} \in \R^k$ (i.e., mean zero risk factor random variable), that is, $f_{t+1} \coloneqq \g + v_{t+1}.$ Then, we specify the observable factor $g_{t+1}$ as potentially linearly correlated with the latent factors through
\begin{equation*}
g_{t+1} = \h v_{t+1} + \e^g_{t+1},
\end{equation*}
where $\h \in \R^{k}$ is an unknown parameter mapping the relation between the latent-factor innovations and the observable factors, and $\e^g_{t+1} \in \R$ is measurement error in $g_{t+1}.$ This specification allows the observable factors to be either simply some component(s) of the true latent factors (e.g., setting $\e^g_{t+1}$ to zero with $\h$ set to $(1,0,0,\dots,0$) or, more generally, some unknown linear function $\h$ of the true latent factors and thus still carry a risk premium. To recover the tradable portfolio representing the nontradable observable risk factor, we map $g_{t+1}$ through $\h$ onto the column space of the true latent factors.

Precisely, the risk premium of an observable factor---our target parameter in this extension---is defined to be the expected excess return of a portfolio with loading (i.e., beta) of 1 with respect to this factor in $g_{t+1}$ and zero loadings on all other factors; in this model, that parameter is $\g_g \coloneqq \h^\top \g.$

\paragraph{Goal} Under the novel asymptotic assumption for this setting of $p,N,T \rightarrow \infty$, we aim to develop estimation procedures for the latent loadings $\G_\b$ and factors $f_{t+1}, \ \forall t;$ in addition, we aim to estimate and conduct inference on $\g_g$ under the novel use of a dynamic latent-factor model and the aforementioned novel high-dimensional asset characteristics.

\subsection{Estimation}\label{s:estimation}

\paragraph{Motivating DSLFM Estimation} The goal is to jointly estimate the loading matrix $\G_\b$ and latent factors $f_{t+1},$ which are not separately identifiable without further restrictions, to be discussed \citep{bai2013principal}. However, to begin, given the model takes the form 
\begin{equation*}
    r_{i,t+1} = z_{i,t}^\top \G_\b f_{t+1} + \e_{i,t+1}
\end{equation*}
where $\e_{i,t+1} = (\e^\b_{i,t})^\top f_{t+1} + \e^r_{i,t+1}$ is the composite idiosyncratic error, we observe that our setting is high-dimensional panel data where we project $\{ r_{i,t+1} \}_{i=1,t=1}^{i=N,t=T}$ onto the column space of $\{ z_{i,t} \}_{i=1}^{N}$ for each time period to estimate each $p$ dimensional time-varying vector $\{ \widehat{\G_\b f_{t+1} } \}_{t=1}^{T}$ where we have to address $p \sim \max(N,T)$ or $p >> \max(N,T).$

Thus, if we utilize the objective function to minimize over the $p-$dimensional choice vector $\G_\b f_{t+1}$ the mean-squared error of $\sum_{i} (r_{i,t+1} - z_{i,t}^\top \G_\b f_{t+1})^2$, we will not only have a noisily estimated design matrix when $p \sim \max(N,T),$ (or, at worst, a nonsingular design matrix when $p>\max(N,T)$) but also a non-convex objective given the interaction between minimization arguments $f_{t+1}$ and $\G_\b$. This rules out implementing low-dimensional (in $p$) methods.

One potential solution would be to introduce sparsity in $z_{i,t},$ given that empirical estimates show, ex post, few covariates contribute the vast majority of the signal, as stated earlier.\footnote{The online implementation of IPCA \cite{kelly2019characteristics} does indeed offer an $\ell_1$ regularization to their MSE objective, which is not discussed in the econometric theory of \cite{kelly2020instrumented}.} This would amount to adding a regularization parameter to the aforementioned objective to combat the curse of dimensionality from $z_{i,t}$ \footnote{One of several ways to interpret the curse of dimensionality is that as the number of covariates increases linearly, the volume of the parameter space to estimate grows nonlinearly; hence the density of the data falls.} However, although potentially helpful for minimizing MSE by decreasing the variance of the estimator, this regularization introduces a bias in estimation, which will lead to invalid asymptotic inference for asset pricing tests, defeating a goal of this work.

We therefore adapt for our purpose the Double Selection Lasso estimator introduced by \cite{belloni2014inference}. The key insight from their work was to introduce an orthogonality wherein, assuming $\G_\b$ is row sparse, the regularization bias from the LASSO first-stage estimation does not pass through to the target parameter of interest when conducting inference.\footnote{The ideas developed in the Double Selection Lasso paper for inference in partially linear models with high-dimensional controls were the basis for the generalization of this idea in the DML procedures as developed in \cite{chernozhukov2018double}. It would likely be closer to the empirical reality to maintain a nonparametric loading \citep{fan2016projected}. As aforementioned, there is thus a natural extension of the work herein to use DML wherein the target variable $c_{t,j}$ is linear while the controls are nonparametrically estimated via a machine learning method, which would require the development of a Neyman Orthogonal score for this panel data setting, perhaps in a similar fashion to \cite{semenova2021debiased}. However, the econometric theory is unknown for inference in the nonparametric setting. Moreover, given the high-dimension asset characteristics, using a conditional independence assumption to obtain a causal parameter may be the most fruitful path toward a causal factor model, a major area of future work (e.g., \cite{lopez2022causal}). We explore the out-of-sample predictive ability of a non-parametric model in the last section of this manuscript.} First, we estimate for each time period $t$ and each characteristic $j$ the scalar $\G_{\b,j}^\top f_{t+1}$ using DSL; then stacking these estimates into a $T \times p$ matrix, we use PCA to obtain separate estimates for latent loadings $\wh{\G}_\b$ and factors $\{ \wh{f}_{t+1} \}_{t=1}^T$; and, finally, soft-threshold $\wh{\G}_\b$ to set the majority of the rows to zero given the assumption of sparsity.

We rewrite the DSLFM model and introduce a first-stage:
\begin{equation}
    \begin{aligned}
    \label{e:model_dslfm}
        r_{i,t+1} &= z_{i,t,j} c_{t+1,j} +  z_{i,t,-j}^\top c_{t+1,-j} + \e_{i,t+1}, & \mathbb{E}[\e_{i,t+1} | z_{i,t}] &= 0, \\
        z_{i,t,j} &= z_{i,t,-j}^\top \d_{t,j} + \epsilon^z_{i,t,j}, & \mathbb{E}[\e^z_{i,t,j} | z_{i,t,-j}] &= 0,
    \end{aligned}  
\end{equation}
where $c_{t+1,j}$ refers to the $j \in \{1,\dots,p \}$ component of $c_{t+1} := \G_{\b} f_{t+1}$ while $-j$ refers to the remaining $p-1$ elements of the vector; $\d_{t,j} \in \R^{(p-1)}$ is an unknown, possibly time-varying, parameter; and, $\e^z_{i,t,j}$ is an unknown scalar random error. $c_{t+1,j}$ is an asset return when its $j-$th characteristic is set to 1 and all other characteristics are set to zero, less its idiosyncratic return $\e_{i,t+1}.$

There are several ways to interpret and justify the first-stage equation as discussed in \cite{belloni2014inference}. Intuitively, the procedure does not rely on perfect model selection for valid inference as instead we not only recover controls $z_{i,t,-j}$ in the second-stage equation for their pricing ability in the cross section of returns but also recover controls with high correlation to our target variable $z_{i,t,j}.$ From a theoretical perspective, the first-stage equation accounts for potential omitted variable bias if one estimated only the second-stage equation. That is, the set of potentially relevant asset covariates is enormous (\cite{chen2020deep} and \cite{bianchi2022dynamics}), and thus a researcher may be motivated to select their preferred subset to ameliorate the curse of dimensionality, which could introduce model selection mistakes. Moreover, it is known LASSO can have poor finite sample model selection performance \citep{chernozhukov2015valid}. Thus, selecting covariates with only the second-stage equation could fail to include relevant controls.

\paragraph{DSLFM Estimation Procedure} Our estimation procedure for $\{ f_{t+1} \}_{t=1}^T$ and $\G_\b$ has three steps: Double Selection Lasso, PCA, and soft-threshold step.
\begin{enumerate}
    \item \textit{DSL}: To estimate $\hat{c}_{t+1,j},$ run $T \times p$ Double Selection Lasso cross-sectional regressions. \footnote{To set the penalty parameter in the LASSO implementations, one can follow the analytic method developed for heteroskedastic, non-Gaussian settings detailed in Appendix A, Algorithm 1 of \cite{belloni2014inference}. For a more modern approach, one can use the bootstrap-after-cross-validation method of \cite{chetverikov2021analytic}. In practice, we use cross validation.} 
    \begin{itemize} 
        \item Run LASSO of $\{r_{i,t+1} \}_{i=1}^N$ on $\{ z_{i,t}\}_{i=1}^N$ for $\hat{c}_{t+1,j}$ and $\hat{c}_{t+1,-j}.$ 
        \begin{itemize}
            \item Let $\hat{I}_1$ denote the nonzero elements of $\hat{c}_{t+1,-j}.$
        \end{itemize}
        \item Run LASSO $\{ z_{i,t,j} \}_{i=1}^N$ on $\{ z_{i,t,-j} \}_{i=1}^N$ for $\hat{\d}_{t,j}.$ 
        \begin{itemize}
            \item Let $\hat{I}_2$ denote the nonzero elements of $\hat{\d}_{t,j}.$
        \end{itemize}
        \item Define the set $\hat{I} = \hat{I}_1 \cup \hat{I}_2 \cup \hat{I}_3$ where $\hat{I}_3$ is the set of controls in $z_{i,t,-j}$ not included in the first two LASSOs that the econometrician thinks are important for ensuring robustness, termed the amelioration set.
        \item Run OLS of $\{r_{i,t+1} \}_{i=1}^N$ on $\{z_{i,t,j}, \tilde{z}_{i,t,-j}\}_{i=1}^N$ where $\tilde{z}_{i,t,-j}$ includes only elements of $z_{i,t,-j}$ in $\hat{I}$. That is,
    \end{itemize}
    \begin{equation*}
    (\hat{c}_{t+1,j}, \hat{c}_{t+1,-j}) \coloneqq \arg \min_{c_{j}, c_{-j}} \{ \mathbb{E}_N [(r_{i,t+1} - z_{i,t,j}c_{t+1,j} - z_{i,t,-j}^\top c_{t+1,-j})] \ : \ c_{t+1,-j,l} = 0, \forall l \notin \hat{I} \}.
    \end{equation*}
    \item \textit{PCA}: To estimate $\G_\b$ and $f_{t+1},$ run PCA on $\hat{C}=\wh{F} \wh{\G}_\b^\top$---a $T \times p$ matrix---to decompose it into $p \times k$ and $T \times k$ matrices $\hat{\G}_\b$ and $\wh{F}.$
    \item \textit{Soft-threshold}: Given the assumed exact row sparsity of $\G_\b,$ we set to zero all rows of $\hat{\G}_\b$ whose row $\ell_1$ norm is below a cross-validated hyperparameter $\lambda$. That is,
    \begin{equation}
        \label{e:soft_threshold}
        \check{\G}_{\b,j} = \left( \norm{ \widehat{\G}_{\b,j} }_1 - \lambda \right)_+ \text{sign} (\norm{ \widehat{\G}_{\b,j} }_1), \ \ j \in \{1,\dots,p\}.
    \end{equation}
\end{enumerate}

This does require running $T \times p$ versions of the cross-sectional Double Selection Lasso regressions, which can be in the thousands in empirical settings; however, these regressions are all computationally light and can be run in parallel. Moreover, this allows for unbalanced panels as each cross-section can have a different number of assets. \footnote{In empirical practice, we find the entire estimation procedure is on the order of ten minutes.} Additionally, these cross-sectional regressions, followed by estimations with the entire panel, mirror the effort of the most commonly used estimation procedure in the factor model setting, namely two-pass Fama-MacBeth regressions.

The high dimensionality of the PCA procedure, given we have a $p \times T$ matrix, is adapted from the existing literature using $N \times T$ matrices \citep{bai2003inferential}. The estimated factor matrix $\widehat{F}$ is the product of $\sqrt{T}$ and the eigenvectors corresponding to the k largest eigenvalues of the $T \times T$ matrix $(Tp)^{-1} \widehat{C} \widehat{C}^\top.$ The estimated factors are normalized such that $\widehat{F}^\top \widehat{F} = I_{k \times k},$ a standard approach. The estimated loadings are $\widehat{\G}_\b = T^{-1} \widehat{C}^\top \widehat{F}.$ We thus see the main challenge in deriving the large-sample theory will be handling the estimation error in using $\widehat{C}$ instead of the unobserved $C.$

The final soft-thresholding step \ref{e:soft_threshold} in our estimation procedure exploits the sparsity in $\G_\b$ to not only reduce the dimensionality of the characteristic space $s$ ($ < < p$) but also remove noise from the characteristics that have low signal-to-noise ratios. \cite{belloni2018high} discuss the general theoretical properties of the soft-threshold estimator with theory-based hyperparameter selection, and its close relation, the better known LASSO and Dantzig selector estimators. We further discuss constraints and selection of the hyperparameter in Appendix \ref{s:proofs}.

\paragraph{Estimating the Risk Premium of an Observable Factor} Under the richer setting that includes the observable factor $g_{t+1},$ our model has an additional specification and moment conditions.
\begin{equation}
    \begin{aligned}
    \label{e:model_ob_risk_premia}
        r_{i,t+1} &= z_{i,t}^\top \G_\b (\g + v_{t+1}) + \epsilon_{i,t+1}, & \E[v_{t+1}] = \E[\epsilon_{i,t+1}] = 0, \ \ \E [v_{t+1} \e_{i,t+1} ] = 0, \\
        g_{t+1} &= \h v_{t+1} + \e^g_{t+1}, & \E[\e^g_{t+1}] = 0, \  \ \E [v_{t+1} \e^g_{t+1}] = 0.
    \end{aligned}
\end{equation}

Our goal is to estimate and conduct inference on $\g_g \coloneqq \h^\top \g.$ Given the latent factors are unobserved, we cannot jointly estimate $v_{t+1}$ and $\h$ without further restrictions. We would have to invoke one of the three classic identification approaches of \cite{bai2013principal}; however, by using the key rotation invariance result of \cite{giglio2021asset}, we can estimate the latent factors up to an invertible rotation matrix $H \in \R^{k \times k}$ and still maintain identification of our target parameter $\g_g.$ That is, both of the underlying parameters will be identified up to this rotation matrix: $\h^\top = \h_0^\top H^{-1}$ and $\g = H \g_0.$ Thus, the target parameter is rotationally invariant to $H:$ $\g_g = \h_0^\top H^{-1} H \g_0 = \h^\top \g.$

For our estimation procedure, we replace the first PCA step of \cite{giglio2021asset} with our above procedure---augmented to use return innovations---to estimate the latent loadings $\check{\G}_\b$ and factor innovations $\hat{v}_{t+1}$ for all $t.$ We then proceed with the latter two steps of the authors' procedure to obtain our target estimator.
\begin{enumerate}
    \item To estimate latent-factor risk premia $\hat{\g},$ run cross-sectional OLS of average returns $\bar{r} \in \R^N$ on averaged estimated latent-factor loadings $\bar{\hat{\b}} = \bar{Z}^\top \hat{\G}_\b \in \R^{N \times k}.$
    \item To estimate latent to observable factor mapping $\hat{\h},$ run a time series OLS regression of $\{ g_{t+1} \}_{t=1}^T$ on factor innovations $\hat{V} \in \R^{T \times k}.$
\end{enumerate}

We can thus form our estimator of the risk premium for the observable factors $g_{t+1}$ by combining these estimators into $\hat{\g}_g = \hat{\h}^\top \hat{\g}.$

This procedure extends the estimation in \cite{giglio2021test} to dynamic loadings and high-dimensional asset characteristics, while inheriting the rotation invariance and the specification consistent with two-pass estimators in this literature. Again, simply applying IPCA instead of PCA in the first step of \cite{giglio2021asset} would not be feasible with $p > \max{N,T}$ or would yield biased inference if an $\ell_1$ penalty were simply added to the IPCA objective. The cross-sectional OLS of average returns on the estimated latent-factor loadings is the standard second step in two-pass Fama-MacBeth regressions, which could be replaced with generalized least squares or weighted least squares to explore asymptotic efficiency gains. The final time series regression is critical to translate the uninterpretable risk premia of latent factors to those of factors proposed by economic theory. Moreover, this procedure handles omitted variable bias which we now briefly discuss.

To illustrate, assume we have a scalar observable factor $g_{t+1},$ which is the first component of a two-dimensional latent-factor innovation vector: $v_{t+1} = (g_{t+1}, v_{2,t+1})^\top$ (i.e. $\h = (1,0)$). The vector-version of our model is thus
\begin{equation*}
    r_{t+1} = z_t \G_{\b,1} (\g_g + g_{t+1}) +  z_t \G_{\b,2} (\g_2 + v_{2,t+1}) + \e_{t+1}.
\end{equation*}

Using the standard Fama-MacBeth two-pass regressions \citep{fama1973risk} will produce bias in estimating $\g_g$ if $v_{2,t+1}$ is omitted. The first step of a time series regression of asset excess returns on $g_{t+1}$ will give a biased estimate of $\hat{\b}_1$ as long as $v_{2,t+1}$ is correlated with both $g_{t+1}$ and $r_{t+1},$ per the standard OVB term: the covariance between between the outcome and the excluded regressor times the covariance between the included and excluded regressor, up to scale. Moreover, in the second step of a cross-sectional regression of average returns on estimated loadings, a second omitted variable bias is introduced if the loading of the omitted factor $\hat{\b}_2$ is correlated with both $\hat{\b}_1$ and $\bar{r}_{t+1}.$ 

Estimating the latent factors via the DSLFM procedure resolves this issue of omitting a potentially relevant factor given one can utilize a consistent estimator of the true number of latent factors, which we assume spans the true factor space.\footnote{The DSLFM could be further extended to estimate the zero-beta rate (i.e., alpha) using a very similar approach to that discussed in Online appendix I.2 of \cite{giglio2021asset}.}

\subsection{Asymptotic Theory}\label{s:asymp_theory}

In this section, we present the asymptotic results for consistent estimation of the latent factors and loadings and the large sample distribution of the nontradable observable factor risk premium estimator under the assumed setting discussed in Section \ref{s:setup} and using estimation procedures discussed in Section \ref{s:estimation} for models \eqref{e:model_dslfm} and \eqref{e:model_ob_risk_premia}. We first provide the regularity conditions sufficient for the validity of the estimation and inference results. For clarity of exposition, we focus on motivating the assumptions and interpreting the results, while theoretical details and mathematical proofs are provided in Appendix \ref{s:proofs}.

Throughout, let $\norm{ A } = [ tr(A^\top A)]^{1/2}$ denote the Frobenius norm of matrix A or $\norm{ x }  = \left( \sum_i x_i^2\right)^{1/2}$ for the $\ell_2$ norm of a vector $x$. Let $\norm{ x }_0 $ and $\norm{ x }_1$ be the usual $\ell_0$ and $\ell_1$ norms, respectively. All limits are simultaneous where we will restrict the rates among $p,T,N$, to allow $p\rightarrow\infty,$ as discussed below.

\subsubsection{Regularity Conditions}\label{s:assumptions}

\paragraph{Consistent Estimators for the Latent-Factor Model} The following assumptions enable the consistent estimation of the factors $\{ f_{t+1} \}_{t=1}^T$ and the loadings $\G_\b$. Let $f_{t+1}^0$ and $\G_\b^0$ be the true factors and loadings such that $f_{t+1} = H f_{t+1}^0$ and $\G_\b = \G_\b^0 H^{-1}$ where $H$ is an unobserved $k \times k$ invertible rotation matrix.

In regard to identification, our results do not require the identification of the true factors $f_{t+1}^0$ and loadings $\G_\b^0$ but rather simply factors (loadings) that span the true factors (loadings) up to the rotation matrix $H.$ \cite{bai2013principal} show identification results for PCA under three different sets of assumptions to pin down the $k \times k$ elements in $H,$ which requires pinning down the covariance matrices of the factor loadings and factors to be diagonal matrices or identity matrices to provide $k(k-1)/2 + k(k+1)/2 = k^2$ restrictions. The researcher can choose which asymptotic covariance matrix to restrict. As we will discuss, we will additionally not need these identification restrictions for the observable factor risk premia given the aforementioned rotation invariance result of the target parameter.

\begin{assumption}[Consistency of DSL]
\label{assump_dsl_consist_main}
\begin{enumerate}
    \item \textit{Bounded Characteristic Portfolios}: For a finite absolute constant $M$ and $\forall t,j,$ $\left|c_{t+1,j}\right| = \left|\G_{\b,j}^\top f_{t+1} \right| < M.$ 
    \item \textit{Sparse Loading}: Loading matrix $\G_\b$ admits an exactly sparse form. That is, for $\exists s \in \mathbb{N}_+, i.e. p > s \geq 1$, $\G_\b$ has at most $s$ nonzero rows:  $\sum_{j=1}^p \mathbb{1} \Bigl\{ \norm{\G_{\b,j} }_1 > 0 \Bigr\} \leq s.$ 
\end{enumerate}
\end{assumption}

These are two critical assumptions for DSL consistency with the additional standard and technical DSL assumptions in Appendix \ref{s:proofs}. Assumption \ref{assump_dsl_consist_main}(i) converts the bounded target parameter, in the traditional DSL context, to the DSLFM context where we require realizations of $c_{t+1,j}$ to be finite-sample bounded by a constant that does not depend on $p,T,N.$ This imposes a bound on the return of characteristic portfolios, that is, the return of a portfolio with characteristic $j$ set to 1 and all other characteristics set to 0. We could instead assume returns are bounded random variable to impose Assumption \ref{assump_dsl_consist_main}(i).

Assumption \ref{assump_dsl_consist_main}(ii) is the key LASSO assumption that the parameter on the control regressors admits an exactly sparse form, which follows from our assumption such that $\forall t,j, \ \norm{c_{t+1,-j} }_0 = \norm{ \G_{\b,-j} f_{t+1} }_0 \leq s$. This sparsity of the loading matrix is supported empirically in asset pricing given the relevance of only a small number of asset characteristics, which we corroborate in our empirical setting. We have thus adapted the classic LASSO sparsity assumption to the empirical reality of cross-sectional asset pricing using high-dimensional asset characteristics. Exact sparsity could be relaxed to approximate sparsity with a similar but alternative high-dimensional econometrics toolkit. 

We next turn to assumptions for consistently estimating the latent factors and loadings. The focus in our work is controlling the estimation error between the infeasible eigendecomposition of $(Tp)^{-1} CC^\top$ and the feasible eigendecomposition of $(Tp)^{-1} \wh{C}\wh{C}^\top,$ given we do not observe $C=F \G_\b^\top$ but instead estimate each element via DSL and then eigendecompose using standard PCA estimators as discussed in Section \ref{s:estimation}.

\begin{assumption}[Consistency of Latent-Factor Model]
\label{assump_latent_fm}
\begin{enumerate}
    \item \textit{Factors}: $\mathbb{E} \norm{ f_{t+1}^{0} }^4 \leq M < \infty$ and $T^{-1} \sum_t f_{t+1}^0 f_{t+1}^{0\top} \rightarrow_p \Sigma_f$ for some $k \times k$ positive definite matrix $\Sigma_f.$
    \item \textit{Factor Loadings}: $\forall j, \ \norm{ \Gamma_{\beta,j} } \leq M < \infty$ and $\norm{ \Gamma_\beta^\top \Gamma_\beta / p - \Sigma_\Gamma } \rightarrow 0$ for some $k \times k$ positive definite matrix $\Sigma_\Gamma.$
    \item \textit{Nonzero and distinct eigenvalues}: from the infeasible eigendecomposition, the $k$ largest eigenvalues $\l_i$ for $i \in \{1, \dots, k\}$ are bounded away from zero. Moreover, the $k$ largest infeasible eigenvalues are distinct, that is, $$\min_{i:i\neq \kappa} |\lambda_\kappa - \lambda_i | > 0.$$
\end{enumerate}
\end{assumption}

Assumptions \ref{assump_latent_fm}(i)-(ii) are standard for factor models where the literature is styled after Assumptions A, B, and C of \cite{bai2003inferential}. Assumption \ref{assump_latent_fm}(i) does not impose i.i.d. factors, as in the classical factor analysis literature, but instead imposes the factors are stationary, strong mixing, and satisfy moment conditions. Assumption \ref{assump_latent_fm}(ii) ensures each latent factor contributes to the second moment of $c_{t+1};$ that is, it imposes all factors are pervasive and excludes weak factors. See \cite{giglio2021test} for adjustments for weak factors. The PCA estimation herein does not require the Assumption C of \cite{bai2003inferential} given our target matrix $C = F \G_\b^\top$ is without an error term; we instead are controlling cross-sectional and temporal dependence using the moment conditions of DSL given in model \eqref{e:model_dslfm} and more technical assumptions in Appendix \ref{s:proofs}.

Assumption \ref{assump_latent_fm}(iii) assumes the $k-$largest eigenvalues from the infeasible and feasible eigendecompositions remain nonzero asymptotically. In finite sample, these are real and nonzero eigenvalues given we are taking the eigendecomposition of a rank $k$ symmetric matrix. It is reasonable to assume we have distinct eigenvalues given, for this not to hold, there would have to be two or more dimensions in the $k-$largest of the $T \times T$ matrix $CC^\top$ that have precisely the same variability.

\paragraph{Estimating Number of Factors} Given the focus of this work is on the consistency of the main estimators and the asymptotic distribution of the risk premium estimator, we assume $k=k^0$ is known.\footnote{The asymptotic distribution of the risk premium estimator is unaffected when the number of factors is estimated because 
\begin{equation*}
    \begin{aligned} 
        \Pr \left( \wh{\g}_g \leq x \right) &= \Pr \left( \wh{\g}_g \leq x, \wh{k} = k^0 \right) + \Pr \left( \wh{\g}_g \leq x, \wh{k} \neq k^0 \right) = \Pr \left( \wh{\g}_g \leq x, \wh{k} = k^0 \right) + o(1) \\  
        &= \Pr \left( \wh{\g}_g \leq x | \wh{k} = k^0 \right) \Pr \left( \wh{k} = k^0 \right)  + o(1) = \Pr \left( \wh{\g}_g \leq x | \wh{k} = k^0 \right) + o(1). 
    \end{aligned}
\end{equation*}} 

\begin{assumption}[Consistent Estimator for Number of True Factors]
\label{assump_number_factors}
For $\bar{k} > k^0,$ let $\wh{k} \coloneqq \arg \min_{0\leq k \leq \bar{k}} IC(k)$ where 
\begin{equation*}
    \begin{aligned} 
        IC(k) &\coloneqq \log (V(k)) + k \left( \frac{p + T}{pT}\right) \log \left( \frac{pT}{p+T} \right) \\
        V(k) &\coloneqq \min_{\G_\b,F} \left(pT\right)^{-1} \sum_{j,T} \left(c_{j,t+t} - \G_{\b,j}^\top f_{t+1} \right)^2.
    \end{aligned}
\end{equation*}
Assume $\wh{k} \rightarrow_p k^0$ without further restriction on the growth rates among $p,T,N$ and $k=k^0$ is known.
\end{assumption}
Assumptions \ref{assump_dsl_consist_main} and \ref{assump_latent_fm} can be shown to be sufficient for consistently estimating, with the above Information Criterion, the number of true factors $k^0$ using the results of Appendix \ref{s:proofs} as in \cite{bai2002determining} and \cite{bai2003inferential}. Although providing this assumption to show the estimator to be studied in simulation, we are instead choosing to impose Assumption \ref{assump_number_factors} in the asymptotic to focus on the main results of this work. Note that commonly used model selection criteria (e.g., AIC or BIC) will not yield consistent estimators, hence the specification above using the contribution of \cite{bai2002determining}.

\paragraph{Inference on Nontradable Observable Factor Risk Premia} The final assumptions are needed to derive the limiting distribution of the risk premium estimator.

\begin{assumption}[Inference]
\label{assump_inf}
There exists a generic absolute constant $M<\infty$ such that for all $p,T,N:$
\begin{enumerate}
    \item \textit{Bounded idiosyncratic errors}: $\E [ ( \sum_t \e_{i,t+1} )^2 ] \leq TM.$
    \item \textit{Bounded scaled factor innovations}: $\E  [ ( \sum_t z_{i,t}^\top \G_\b^0 v_{t+1}^0 )^2 ] \leq sTM.$
    \item \textit{Bounded measurement errors}: $\E [ ( \epsilon^g_{t+1} )^2 ] \leq M.$
    \item \textit{Convergence of characteristics}: $\frac{1}{NT} \sum_i \sum_{t'} \E[z_{i,t,j}] z_{i,t',j'} \rightarrow_p \mathcal{Z}_{t,j,j'}$ uniformly over $t,j,j'$ for $j,j' \in \{1,2,\dots,p\}$ and a nonstochastic finite constant $\mathcal{Z}_{t,j,j'} \in \R.$
    \item \textit{CLT}: As $T\rightarrow \infty,$ the following joint central limit theorem holds: 
    $$ \frac{\sqrt{T}}{T} \sum_t \left( 
    \begin{matrix} 
    v_{t+1}^0 \epsilon_{t+1}^g \\ 
    \Pi_t v_{t+1}^0 
    \end{matrix} \right) \xrightarrow{d} \mathcal{N} (0, \Phi)$$ 
    where random matrix $\Pi_t \in \R^{k \times k}$ and nonstochastic matrix $\Phi \in \R^{2k \times 2k}$ are defined in Appendix \ref{s:proofs}.
\end{enumerate}
\end{assumption}

Assumption \ref{assump_inf}(i) bounds the second contemporaneous and cross-moments of the idiosyncratic errors, aligning with the time and cross-section dependence assumptions of \cite{bai2003inferential} Assumption C. The assumption would hold if we assumed $\e_{i,t+1}$ are uncorrelated across $t$, which is a simplified yet plausible assumption given the low signal-to-noise environment of asset pricing. We have thus relaxed the temporal dependence to the specified rate $T$.

Assumption \ref{assump_inf}(ii) bounds the squared time series average of the factor innovations scaled by the factor loadings. In the static factor model context of \cite{giglio2021asset}, this holds in large sample by a simple LLN argument given the static loadings are not a function of $t$ and the factor innovations are mean zero random variables. Thus, we are ensuring the $\G_\b^0$ selected columns of $Z_t$ keep the scaled $v_{t+1}^0$ sufficiently small.

Assumption \ref{assump_inf}(iii) bounds the second moment of the observable factor measurement errors for use in proving $\norm{ \e^g } = O_p (\sqrt{T}).$ It is not a stringent assumption because we are bounding a zero mean scalar random variable. This is nearly an identical assumption and usage to \cite{giglio2021asset} Assumption A8.

Assumption \ref{assump_inf}(iv) provides a convergence result such that the squared first moment for two different characteristics averaged over time and across assets is a nonstochastic finite constant. This is a weaker assumption on the distribution of characteristics than the DSL moment conditions discussed in Appendix \ref{s:proofs}.

Assumption \ref{assump_inf}(v) is the assumed central limit theorem for the $2k$ (low) dimensional mean zero random variable given the models' \ref{e:model_dslfm} and \ref{e:model_ob_risk_premia} moment assumptions, which is satisfied by various mixing processes. The second moments of the later $2k$ random variables are bound already in Assumptions \ref{assump_inf}(i)-(ii). We nevertheless directly assume the needed CLT. This extends for our inference result the assumed CLT at the same rate in Assumption F4 of \cite{bai2003inferential} and the assumed CLT at the same rate in Assumption A11 of \cite{giglio2021asset}. Note that although we have the same two mean zero random vectors, our factor innovations are scaled by $\Pi_t$ instead of a constant $1$ given the dynamic factor loadings of our model.

\subsubsection{Theory Results}\label{s:asymp_results}

This section presents the three main theoretical results.

\paragraph{Consistent Estimators for the Latent-Factor Model} We present the first two results showing the consistency of the latent-factor model estimators.

\begin{proposition}[Consistency of Latent Factors]
Under the model \eqref{e:model_dslfm}, Assumptions \ref{assump_dsl_consist_main}, \ref{assump_latent_fm}, \ref{assump_number_factors}, and DSL Assumptions in Appendix \ref{s:proofs}.2 where $T,N,p \rightarrow \infty,$ then for all $t$ the latent-factor estimator described above has the property that
$$  \wh{f}_{t+1} - H^\top f^0_{t+1}  = O_p \left( \sqrt{\frac{s \log(Tp)}{N}} \right).$$
\end{proposition}

The proof is in Appendix \ref{s:proofs}. This result establishes the convergence rate of the latent factor estimator in a dynamic latent-factor model with high-dimensional characteristics. If the factor loadings were static and known, $\beta_i^0$ for all $i,$ then, $f_{t+1}^0$ would be estimated via a cross-sectional least squares with a convergence rate of $\sqrt{N}.$ \cite{bai2003inferential} establishes in Theorem 1(ii), under $N,T\rightarrow \infty$ for a static latent-factor model, the foundational result of a convergence rate of $\min \left( \sqrt{N}, T \right)$ for the consistency of the latent-factor estimator for the rotated true factors $H^\top f_{t+1}^0.$ Incorporating dynamic loadings parameterized by high dimensional characteristics comes at the cost of slowing the rate to $\sqrt{pN / s \log(Tp)},$ which is nevertheless still reasonable for typical values of $p,T,N.$ Additional standard DSL rates, which are less restrictive, are in Appendix \ref{s:proofs}.

Our rate is primarily driven by the $\sqrt{N / \log(Tp) }$ rate uniform consistency over $t$ and $j$ of the DSL estimation error $| \wh{c}_{t+1,j} - c_{t+1},j |$ as shown in Lemma A1 in Appendix \ref{s:proofs}. Given the model for $C = F^0 \G_\b^{0 \top}$ contains no error, the eigendecomposition of the unobserved $C$ is exact for $F^0 H$ as shown in Lemma A6; and, thus, the estimation error from using $\wh{C}$ instead of $C$ drives this first main result. The assumed sparsity in $\G_\b^0$ does improve the rate with the $p / s$ ratio.

It is worth reiterating that under our setting $F^0$ and $\G^0_\b$ are not separately identifiable, hence the $k \times k$ invertible matrix transformation $H$ appears in each asymptotic result. Similarly, $\wh{F} \wh{\G}_\b^\top$ is an estimator of the identifiable, rotation invariant common component $C,$ which is corroborated by simulation results. Moreover, in many cases knowing $F^0 H$ is equivalent to knowing $F^0;$ for example, the regressor $F^0$ will give the same predicted values as using $F^0 H$ as a regressor given they have the same column space.

\begin{proposition}[Consistency of Latent-Factor Loadings]
Under the model \eqref{e:model_dslfm}, Assumptions \ref{assump_dsl_consist_main}, \ref{assump_latent_fm}, \ref{assump_number_factors}, and DSL Assumptions in Appendix \ref{s:proofs}.2 where $T,N,p \rightarrow \infty,$ then the latent loading estimator described above has the property that
$$\left( \check{\G}_{\b} - \G_{\b}^0 H^{-1} \right) = O_p \left( \sqrt{\frac{s \log(Tp)}{N}}  \right).$$
\end{proposition}

The proof is in Appendix \ref{s:proofs}. This result establishes the convergence rate of the latent loading estimator in a dynamic latent-factor model with high-dimensional characteristics. When the factors, $f_{t+1}^0,$ for all $t,$ are observable, static loadings $\beta_i^0$ can be estimated by a time series regression with a convergence rate of $\sqrt{T}.$ \cite{bai2003inferential} establishes in Theorem 2(ii), under $N,T\rightarrow \infty$ for a static latent-factor model, the foundational result of a convergence rate of $\min \left( N, \sqrt{T} \right)$ for the consistency of the latent-factor loading estimator $\wh{\b}_i$ for the rotated true factor loadings $H^{-1} \b_i^0.$ Incorporating dynamic loadings parameterized by high dimensional characteristics comes at the cost of slowing the rate to $\sqrt{N / s \log(Tp)},$ which is nevertheless still reasonable for typical values of $p,T,N.$

The rate follows similar reasoning to that of the latent-factor estimator in Proposition 1. However, here we are presenting the rate of the final soft-threshold estimator---derived using recent results in high dimensional econometrics \cite{belloni2018high}---wherein we use the uniform estimation error between the eigendecomposition of $\wh{C}$ for $\wh{\b}_{\b,j}$ and the infeasible loading $\wt{\b}_{\b,j}$ from decomposing the unobserved $C,$ which eliminates $p$ in our rate from Proposition 1. The $\sqrt{N / log (Tp) }$ is similarly driven by the uniform consistency over $t$ and $j$ of the DSL estimation error $| \wh{c}_{t+1,j} - c_{t+1},j |,$ which is the key result used to establish these consistency propositions along with typical high dimensional random matrix theory (e.g., Davis Kahan Theorem, Weyl Inequality, and recent tools in high dimensional econometric theory found in \cite{belloni2018high}).

\paragraph{Inference on Nontradable Observable Factor Risk Premium} Finally, we present the asymptotic normality of the nontradable observable factor risk premium estimator.

\begin{theorem}[Normality of Observable Factor Risk Premium]
Under the models \eqref{e:model_dslfm} and \eqref{e:model_ob_risk_premia}; Assumptions \ref{assump_dsl_consist_main}, \ref{assump_latent_fm}, \ref{assump_number_factors}, \ref{assump_inf}; DSL Assumptions in Appendix \ref{s:proofs}.2; and if $T s^2 \log (Tp) / N  \rightarrow 0,$ then as $T,N,p \rightarrow \infty$ the estimator $\hat{\g}_{g}$ obeys
$$\sqrt{T} \frac{(\hat{\g}_{g} - \g_{g})}{\s_{g}} \xrightarrow{d} \Nc (0,1),$$
where $\s_{g}$ is defined in Appendix \ref{s:proofs}.
\end{theorem}

The proof is in Appendix \ref{s:proofs}. This result establishes $\sqrt{T}$ asymptotic normality of the nontradable observable factor risk premium estimator from \cite{giglio2021asset} extended to the setting of dynamic factor loadings with high dimensionality characteristics. At a high level, our proof follows a similar approach yielding, as seen in the proof of Theorem 1, the same two asymptotically nonnegligible terms as in \cite{giglio2021asset}. The first term arises from the time-series regression of the observed factor on the latent loadings where again the latent loading estimation error is higher order. The latter term in the $2k$ random vector in Assumption \ref{assump_inf}(v) arises from the cross-sectional regression of averaged asset excess returns on averaged factor loadings, where the factor loading estimation error and idiosyncratic error term are higher order. Although \cite{giglio2021asset} have this same second term, ours is more complicated given the dynamic loadings, which necessitates the convergence Assumption \ref{assump_inf}(iv). A direct application of the delta method on the sum of these two terms yields the result in Theorem 1. 

The crucial rate assumption is $T s^2 \log (Tp) / N  \rightarrow 0,$ which controls the estimation error for the unobserved averaged latent-factor loadings $T^{-1} \sum_t \b_{i,t}.$ This is similar to \cite{bai2003inferential} and \cite{giglio2021asset}, which require $T/N \rightarrow 0$ to use the estimated factors or loadings as generated regressors. However, we have slowed the rate again due to the high-dimensionality in $p.$ This is our slowest required rate.

Given the rotation invariance of the target parameter $\g_g^0$, the unobserved rotation matrix $H$ does not appear in the asymptotic distribution, in contrast to the consistency results. In finite sample, we corroborate this result in the to-be-discussed simulations. In a similar vein, it warrants noting the asymptotic efficiency loss due to not observing the factors and loadings could be large when $N$ is relatively small. Also, we have assumed we directly observe the number of tree factors $k^0,$ which would require estimation in practice and thus likely contribute estimation error to affect finite sample performance.

In simulation, we use the plug-in estimator for $\s_g$, which has satisfactory finite-sample coverage properties. However, one can establish a consistent variance estimator using a \cite{newey1987hypothesis} style plug-in estimator of the asymptotic variance $\s_g$ with lag corrections to account for temporal dependence as in \cite{giglio2021asset} Section IV Part E.

\subsection{Asset Pricing Tests}\label{s:asset_pricing_tests}

In this section we develop three tests central to our empirical analysis. The first uses the asymptotic normality of the observable factor risk premium for a statistical test of nonzero risk compensation. The second statistic informs the incremental significance of any specific asset characteristic. The third and final discusses how we empirically measure whether the DSLFM contributes predictive signal above and beyond a random walk.

\paragraph{Testing Nontradable Observable Factor Risk Premium} An empirical application of the DSLFM model will address whether a nontradable observable factor, namely, inflation, carries a nonzero risk premium in the crypto asset class. The target parameter $\g_g$ captures the risk premium of the (inflation) factor-mimicking portfolio within the crypto asset class as recovered by the estimated dynamic latent-factor model. We are interested not only in the sign of the parameter, but also, in practical settings, in whether a confidence interval suggests a risk premium of economic significance.

We test the hypothesis $H_0: \g_g = 0 \ \ \ \text{vs.} \ \ \ H_1: \g_g \neq 0$ using the risk premium estimation procedure described in Section \ref{s:estimation} with a plug-in variance estimator $\wh{\s}_g$ for $\s_g.$ Given the asymptotic normality of Theorem 1, we form a confidence interval 
$$\g_g \in \left[\wh{\g}_g - c(1-\alpha/2) \wh{\s}_g, \wh{\g}_g + c(1-\alpha/2) \wh{\s}_g \right]$$
where the critical value $c(1-\alpha/2)$ is the $1-\alpha/2$ quantile of a $N(0,1)$ distribution for the researcher-specified level of the test $\alpha.$ We find in the coming simulation acceptable finite sample coverage for this confidence interval.

\paragraph{Testing Characteristic Significance} The large-sample distribution of the latent loading is unknown given the DSL regularization. Even inference in simple cross-sectional LASSO is complicated \citep{lee2016exact}. Instead, we develop a simple bootstrap procedure to infer whether a specific characteristic significantly contributes to loading $\G_{\b}$. We leave for subsequent research developing the supporting theory of this bootstrap procedure or develop the asymptotic distribution of a consistent latent loading estimator in this setting. We test the hypotheses $\forall j$
\begin{equation*}
H_0: \G_\b^\top = \left[ \G_{\b,1}, \dots, \G_{\b,j-1}, 0, \G_{\b,j+1}, \dots, \G_{\b,p} \right] \ \ \ \text{vs.} \ \ \ H_1: \G_\b^\top = \left[ \G_{\b,1}, \dots, \G_{\b,p} \right].
\end{equation*}

That is, we ask whether characteristic $j$ contributes to the factor loading through $k \times 1$ mapping vector $\G_{\b,j}.$ This allows the researcher, using a large number of characteristics, to systematically ask what characteristics contribute to the latent-factor model, instead of an ad hoc selection. We thus set the entire $k \times 1$ vector to zero so the characteristic contributes to predicting the variation in returns through none of the $k$ factors.

Our procedure is to test the alternative hypothesis model, with the unconstrained characteristic $j,$ and then form the test statistic
$$W_{\G,j} = \G_{\b,j}^\top \G_{\b,j}.$$
Using bootstrapped standard errors, we assess whether this test $W_{\G,j}$ statistic is statistically distinguishable from zero. 

\paragraph{Testing Out-of-Sample Performance} To study the out-of-sample pricing ability of the DSLFM, we use the ``predictive $R^2$'' defined as

$$\textrm{Predictive} \ R^2 = \frac{\sum_{i,t} \left(r_{i,t+1} - z_{i,t}^\top \Check{\G}_\b \wh{\lambda}_t \right)^2}{\sum_{i,t} r_{i,t+1}^2}$$
where $\wh{\lambda}$ is the moving average of the estimated factors in previous time periods over a cross-validated window size. This measure captures whether the model forecasts realized returns better than a random walk; or, said differently, it represents the fraction of realized return variation explained by the model's description of expected returns through exposure to systematic risk. This specification allows the model's estimated conditional expected returns to be driven not just by the dynamic factor loadings, estimated using high-dimensional asset characteristics, but also by time-varying risk prices $\lambda_t.$

\subsection{Simulations}

This section presents a brief study of the finite-sample performance of the dynamic latent-factor model estimators and coverage properties of the inference procedure using Monte Carlo simulations. To summarize, we find the estimation errors for factors and loadings are comparable to IPCA and the Three Pass estimator of \cite{giglio2021asset} in low-dimensional settings while superior in high-dimensional settings. This holds even in rather small samples with low signal to noise ratios, reflecting the empirical reality of cross-sectional asset pricing. Moreover, we find estimation errors and coverage properties for the observable-factor risk premium to be comparable to \cite{giglio2021asset} in low-dimensional settings while superior in high-dimensional settings. We now present the design, followed by the results.

\subsubsection{Simulation Design} 

First, we describe the data-generating process for given $N,T,k$ where we follow the finite sample simulation study of IPCA \citep{kelly2020instrumented}. That is, the DGP is favorable to IPCA. We calibrated the simulated data to parameter estimates from IPCA fit to our weekly panel of crypto asset data using all sixty three asset characteristics.

Latent factors $f_{t+1}$ are simulated from a $VAR(1)$ model employing normal innovations that was fit to the estimated IPCA factors. Asset characteristics are simulated from a $p$ variable panel $VAR(1)$ model with normal innovations, which was fit to the demeaned empirical weekly panel of randomly selected, without replacement, $p$ asset characteristics. For each asset, we set the means of the characteristics to a bootstrap sample from the empirical distribution of time series asset characteristic means. The idiosyncratic error $\e_{i,t+1}$ is simulated from an i.i.d. normal distribution whose variance is calibrated such that the population $R^2$ of the model is approximately 20\%, matching the empirically estimated value from fitting IPCA. The measurement error $\e^g_{t+1}$ is simulated in a simple fashion but the $R^2=1-\E [e^g]/\E[g]$ is calibrated to approximately 40\%. $\eta = (1,0,\dots,0)$ and the loadings $\G_\b$ are set to the empirically estimated values where $p-s$ rows are set to zero where $s=p/10.$ Finally, observable factors and returns are generated according to models \eqref{e:model_dslfm} and \eqref{e:model_ob_risk_premia}.

The simulation studies results across $S=200$ Monte Carlo draws. Hyperparameters are fixed at $N=500, \ T=100,$ and $k=3.$ To compare the performance of estimators under low-dimensional and high-dimensional characteristics, results are generated for $p=10$ and $p=50.$ We report results for a variety of estimators, including latent loadings $\G_\b$, latent factors $F,$ average factor loadings $\bar{\b},$ latent matrix $C=F \G_\b^\top,$ and observable factor risk premium $\g_g.$

The benchmark estimation and inference procedures are IPCA and the three-pass estimator of \cite{giglio2021asset}, given DSLFM's basis on these foundational models.\footnote{Many thanks to Matthias Buechner and Leland Bybee for the IPCA implementation \url{https://github.com/bkelly-lab/ipca}.} We focus on two comparisons: first, the estimation error of theoretically consistent latent loading $\G_\b$ and latent factor $\{ f_{t+1} \}_{t=1}^T$ IPCA and DSLFM estimators; and, second, coverage properties of the observable factor risk premium estimator. We do study estimation errors for additional estimands as relevant (e.g., the three-pass estimator does not estimate latent loadings $\G_\b$ nor latent matrix $C$, while IPCA does not have an observable factor risk premium estimation nor inference procedure). 

\subsubsection{Simulation Results}

Table \ref{f:sim} reports results. In the low-dimensional setting of $N=500,T=100,p=10,s=1,$ we find DSLFM to obtain smaller estimation errors for $\G_\b$ as compared to IPCA; however, IPCA has an order of magnitude lower estimation errors for $F.$ DSLFM's outperformance for the latent loading is driven by taking on higher bias yet substantially lower variance of the estimator; this is obtained from soft-thresholding many of the rows. In comparing other auxiliary estimands, DSLFM obtains lower estimation error for the time-series averaged factor loadings $\bar{\b}$ although higher error for the latent matrix $C = F \G_\b^\top.$ We attribute these results to IPCA fitting data simulated to match the fits of an empirically estimated IPCA model, yet we employ an exact row-sparsity structure in the true latent loadings $\G_\b.$

DSLFM slightly under-covers the 90\% and 95\% confidence intervals in the low-dimensional setting. The three-pass estimator obtains similar estimation error for the target parameter of the observable factor risk premium, but has finite-sample intervals that slightly over-cover. 

Moving to the high-dimensional setting of $p=50,$ we find DSLFM to again obtain smaller estimation errors for latent loadings, yet now the estimation errors for the latent factors are of the same order as IPCA. In both cases, DSLFM takes on bias from its regularization methods, although DSLFM's latent factor estimator's variance is still higher than IPCA. DSLFM is now an order of magnitude improvement for the average factor loadings and a factor of two improvement for the latent matrix $C.$ Finally, coverage proprieties of the risk premium estimand for both the three-pass estimator and DSLFM estimators are degraded under high-dimensionality.

We hope to add even higher dimensional results with hyperparameters closer to the empirical values. Do note, given the DSLFM's large sample theory, it is constrained by $N > T,$ which although is the case for the panel of crypto asset returns, this is not the case for all cross-sectional asset pricing settings. Moreover, the DSLFM's performance was boosted by the assumed exact sparsity in the latent loadings; we hope to add results for approximate sparsity, which is likely closer to the empirical reality. Nevertheless, the DSLFM performs well as compared to state-of-the-art benchmark methods, especially under the setting for which it was developed: high-dimensional asset characteristics.

\section{Empirical Applications}\label{s:empapps} In this section, we first establish the pricing ability of a variety of benchmark factor models and compare these results to those of the DSLFM. With the DSLFM, we then utilize the asset pricing tests developed in Section \ref{s:asset_pricing_tests} to elucidate the drivers of returns and conduct inference for crypto's inflation risk premium.

See XYZ for the description of the data, which is the same panel we use in this paper. Tables \ref{f:char_desc_stat_panela} and \ref{f:char_desc_stat_panelb} report descriptive statistics for the panel's dependent variable, asset excess returns over the subsequent week, and the set of sixty three asset characteristics.

\paragraph{Multivariate Observable Factor Models} We now turn to the out of sample performance of low dimensional factor models in estimating risk premia of crypto asset excess returns, beginning with multivariate observable factor models. Instead of selecting a small number of observable factors based on individual univariate performance, we instead form, using the sixty three characteristics, all combinations of one-, two-, and three-factor models to select the best model of each size based on its performance in combined period of the second half of 2021 and first half of 2022.

In detail, to select multivariate factor models, we perform the following procedure. First, we form strictly time varying risk factors using each of the sixty three asset characteristics as the top minus bottom value-weighted quintile portfolio excess returns. We thus do not normalize characteristics. To form predicted asset returns, we next estimate each asset's static factor loading as the contemporaneous in sample time series regression of excess returns on factor(s); estimate risk factor conditional means as the time series average of the factor in sample; and, predict the asset's return for the next week using the dot product. We use this procedure to fit models in 2018 through the first half of 2021 and, with an expanding window, predict week by week for the combined validation period of the second half of 2021 and the first half of 2022. For all one-factor models, the sixty three choose two two-factor models, and the sixty three choose three three-factor models, we then select the best model of each size based on the predictive $R^2$ for the fifty two weeks in the validation period. Throughout, we have to reform the panel each month for the relevant included assets. To compare to the literature, we also formed the \cite{liu2022common} Fama-French style three-factor model using their CMKT, CSMB, and CMOM risk factors.\footnote{Code to replicate can be found at \url{https://github.com/adambaybutt/crypto_asset_pricing/blob/main/code/10a-low_dim_fm_multi.ipynb}.}

Table \ref{f:low_dim} presents results for the test period, i.e., the second half of 2022. The best multivariate observable factor models were size; illiquidity and size; and, size, one month momentum, and three month volatility. Interestingly, the model selection process incorporated size in all three models. The predictive $R^2$ for all three models and the benchmark model were all negative, performing worse in MSE pricing ability than a random walk. However, although statistically insignificant, all three multivariate observable factor models had economically significant weekly time series average excess return spreads of about 1\% with associated Sharpe ratios of 1.31-1.72, which all beat the benchmark model with a return spread of 0.5\% and a Sharpe ratio of 0.65. All four associated alphas show these long-short strategy returns are meaningfully uncorrelated. The illiquidity and size two-factor model achieved the highest Sharpe of 1.72, which is suggestive evidence of a low number of factors being optimal. Although the strategies replicated out of sample, we should note these results are again before transaction costs and are for a very small test period.

\paragraph{Static Latent-Factor Model} We next study the out of sample performance of a static latent-factor model in estimating risk premia of crypto asset excess returns. We are not only interested in how learning the factors from the data changes the out of sample pricing ability, but also in developing benchmarks for the high-dimensional dynamic latent-factor model. We use the classic approach of PCA \citep{bai2003inferential} to estimate one- through five-latent factor models.

In detail, for each month in the test period (i.e., the second half of 2022), we form factor(s) using the matrix of contemporaneous excess asset returns for the relevant included assets; for each asset, we run a time series predictive regression of its weekly excess returns on the factor(s) to obtain its factor loading(s); and, we use each asset's factor loadings and the PCA-estimated factors to generate predicted returns for the out of sample month.\footnote{Code to replicate can be found at \url{https://github.com/adambaybutt/crypto_asset_pricing/blob/main/code/10b-low_dim_fm_pca.ipynb}.}

Table \ref{f:low_dim} reports results for the test period, i.e., the second half of 2022. The predictive $R^2$ for all five latent-factor models were all negative, performing worse in MSE pricing ability than a random walk. Although all five models had positive return spreads, only three of the five models had Sharpe ratios---0.78, 1.24, and 1.34---in the range of the observable factor models; however, there was not a clear pattern across the number of latent factors. This is suggestive evidence of latent factors not offering a clear benefit over the multivariate observable factor models. Their out-of-sample performance yielding uniformly positive Sharpe, generated without using asset characteristics, maintains a benchmark for the richer models.

\paragraph{Dynamic Latent-Factor Model with Low-Dimensional Characteristics} We close our study of the out of sample performance of low dimensional factor models in estimating risk premia of crypto asset excess returns. We investigate the performance of IPCA, a dynamic latent-factor model where the number of asset characteristics must be smaller than the number of assets and time periods.\footnote{We are grateful to Matthias Buechner and Leland Bybee for the IPCA implementation at \url{https://github.com/bkelly-lab/ipca}.} Given the small number of assets in the panel (i.e., there are less than two dozen for the majority of the weeks), we have to, outside of IPCA, select features from the sixty three asset characteristics. We chose to just use the characteristics listed in Table \ref{f:uni_factors_sig}.\footnote{We realized after doing this that this biases IPCA favorably as the univariate results used the IPCA test period, which is one of several reasons that we have not even formed 2023 data to repeat our out-of-sample exercises in fresh and larger data.} We again reform the panel each month and normalize period-by-period features to linearly spaced on $[0,1]$. Finally, we do not specify a constant to allow for mispricing effects but rather explain variation in expected returns using exposure to common latent risk factors. \footnote{Code to replicate our results can be found at \url{https://github.com/adambaybutt/crypto_asset_pricing/blob/main/code/10c-low_dim_fm_ipca.ipynb}.}

Table \ref{f:low_dim} reports results for the test period, i.e., the second half of 2022. Predictive $R^2$ are positive except for a five factor specification with the maximum predictive $R^2$ of 0.18\% for the three-factor model. All five models have economically significant weekly excess return spreads, from 1.5\% to 3.1\%, and associated annualized Sharpe ratios, from 2.07 to 4.07. The one- through four-factor models have statistically significant time-series average weekly excess return spreads for the zero-investment long-short strategies of, respectively, 2.8\%, 2.9\%, 3.1\%, and 2.4\%. The alphas remain statistically and economically significant with little return lost to the market. Remarkably, there are only two quintile portfolios out of twenty five that break monotonicity. Sharpe ratios nearly monotonically decline with the number of factors; the three factor model edges out the two factor model by a difference of 0.13. Although again these results do not account for transaction costs and the test period is short, the dynamic latent-factor model estimated with IPCA dominates the static factor models. 

\paragraph{Dynamic Latent-Factor Model with High-Dimensional Characteristics} We study three questions using the DSLFM. We first compare the out of sample predictability in the same test period to that of the previous factor models, in addition to understanding the characteristics driving returns and to estimating the inflation risk premium in the crypto asset class. The setting is the same weekly panel, reformed each month with the relevant assets, with asset characteristics normalized to 0 to 1.\footnote{Code to replicate can be found at \url{https://github.com/adambaybutt/crypto_asset_pricing/blob/main/code/12-dslfm.ipynb}.}

The DSLFM estimation procedure is outlined in Section \ref{s:estimation} and the test procedures are defined in Section \ref{s:asset_pricing_tests}. There are several hyperparameters for the statistician to chose, for which we are empirically motivated to use cross validation. Specifically, we generate predicted returns week by week in the same validation period of the second half of 2021 through the first half of 2022 by using an expanding window training data set using all previous weeks from the start of the panel. For models with one to five latent-factors, we cross validate the relevant hyperparameters, including the soft thresholding hyperparameter, the lasso penalty parameter, and the number of trailing weeks to average fitted latent factors over to form predicted factors. We present results for models with the best predictive $R^2$ in the validation period.

Table \ref{f:dslfm_oos} presents out of sample test period results for the DSLFM. Only one out of the five models had a positive predictive $R^2.$ Eight out of ten models, when forming equal-weighted and value-weighted portfolios within quintile, had economically significant time series average returns for the long-short strategies, which were maintained when studying the associated alphas. The equal-weighted portfolios had in all but one case, the five latent-factor specification, superior Sharpe ratios, driven by both improved return spreads and lower volatility.

Given the small panel, the poor pricing ability of the DSLFM with more factors is perhaps not surprising given it could be over parameterized and noisily estimated. The factor loading matrix grows by $p$ with each additional specified latent factor. Nevertheless, the large majority of the long-short strategies obtained economically significant Sharpe ratios and associated alphas, representing an improvement over the observable factor models and PCA. However, IPCA outperformed, which is perhaps driven by a meaningful signal to noise ratio improvement through the feature selection done before fitting IPCA. We will explore this in 2023 data, which will yield much more data given the wide cross-section relative to preceding years.

Table \ref{f:dslfm_char_imp} presents bootstrapped results on asset characteristic importance to understand the drivers of returns. Exchange inflows and outflows were the two statistically significant characteristics with point estimates on their importance more than an order of magnitude larger than the next characteristic. Again, we observe the importance of onchain data. This empirically supports approximate sparsity as a reasonable assumption, given the fast decay with a long tail on the importance of these characteristics. Our theory, although it would accommodate approximate, assumes exact sparsity for simplicity. Interestingly, none of the statistically significant univariate factor strategies were significant in the DSLFM characteristic importance. However, all six were at least in the top half, and, in practice, the importance of studying exchange flows is well known. In future work, we will compare these results to the importance measures available in IPCA.

Finally, to demonstrate the extensibility of the DSLFM, we conduct inference on an observable factor risk premium, namely testing for a nonzero premium for ten year expected inflation risk within the crypto asset class. This has been a long-standing research question to understand the relationship between crypto's returns and inflation. Early proponents of Bitcoin and other cryptocurrencies framed these as an outside option or hedge against traditional fiat currencies. To study this question, we use our extended model with one factor and the associated estimation procedure---as described in Section \ref{s:estimation} in how we extend \cite{giglio2021asset}---to recover the 10-year expected inflation mimicking portfolio and measure its risk premium. 

The inflation risk premium was estimated to be a statistically significant 1.4 bps with a standard error of 0.0097 bps. This translates to a 7.3\% annual excess return, suggestive of positive compensation for investors holding an inflation-hedged crypto portfolio, ceteris paribus. The result corroborates similar findings using more simple methods, detailed in Section \ref{s:empfacts}, with a dynamic latent factor model with superior pricing ability. One could attribute this to several aspects of the DSLFM, for example, it allows for regime changes with time-varying loadings, it incorporates rich structure with the full asset characteristics, among other reasons. There are limitations however, including the slow asymptotic rate with this inference procedure, as discussed in Section \ref{s:asymp_theory}, which is exacerbated by the small cross-section in our setting. Thus, although significant, we should interpret this result as suggestive and seek replication. 

\section{References} 

\bibliography{\bib}

\newpage
\appendix

\newpage
\section{Technical Details and Proofs}\label{s:proofs}

\subsection{Notation}

Let $\mathbb{E}_N [x_i ] \coloneqq N^{-1} \sum_t x_i$ for random variables $\{ x_i \}_{i=1}^N.$ 

Let $\mathbb{I}_k$ be a $k\times k$ identify matrix. Let $\norm{ \cdot }$ be the Frobenius norm for a matrix and the $\ell_2$ norm for a vector; $\norm{ \cdot }_1$ be the $l_1$-norm; $\norm{ \cdot }_2$ be the spectral norm for a matrix; and, $\norm{ \cdot }_\infty$ be the maximum element of the matrix or vector. Let $a \lor b = max(a,b).$ We also use the notation $a \lesssim_P b$ to denote $a = O_p (b)$ for $a,b \in \mathbb{R}.$

Define the following random variables: $r_{t+1} = (r_{1,t+1}, \dots, r_{N,t+1})^\top \in \mathbb{R}^N;$ \\ $z_{t,j} = (z_{1,t,j}, \dots, z_{N,t,j})^\top \in \mathbb{R}^{N};$ $Z_{t,-j} = (z_{1,t,-j}, \dots, z_{N,t,-j})^\top \in \mathbb{R}^{N\times (p-1)};$ \\ $\epsilon_{t+1} = (\epsilon_{1,t+1}, \dots, \epsilon_{N,t+1})^\top \in \mathbb{R}^N;$ $\epsilon^z_{t,j} = (\epsilon^z_{1,t,j}, \dots, \epsilon^z_{N,t,j})^\top \in \mathbb{R}^N,$ and so on.

For $A \subset \{ 1, \dots, p \},$ let $Z_{t,-j} [A]$ denote the subset of the columns of $Z_{t,-j}$ that are elements of the set $A.$ Let $\mathcal{P}_A \coloneqq Z_{t,-j} [A] \left(Z_{t,-j} [A]^\top Z_{t,-j} [A] \right)^{-1} Z_{t,-j} [A]^\top $ be the projection operator that maps vectors in $\mathbb{R}^N$ into $\mathrm{span} (Z_{t,-j} [A]).$ Let $\mathcal{M}_A = \mathbb{I}_N - \mathcal{P}_A$ be the operator that projects vectors in $\mathbb{R}^N$ into the subspace orthogonal to $\mathrm{span} (Z_{t,-j} [A]).$

\subsection{Consistency of Double Selection Lasso}

We provided two critical Double Selection Lasso (DSL) assumptions in Assumption \ref{assump_dsl_consist_main}, to which we add the following standard DSL assumptions, adapted to the DSLFM setting. Let there exist absolute sequences $\delta_{N,T} \searrow 0$ and $\Delta_{N,T} \searrow 0.$

\begin{assumption}[ASR: Approximate Sparse Regressors]
\label{ASR} $ $
\begin{enumerate}
    \item \text{Sparsity of Confounding}: The confounding mapping $\delta_{t,-j}$ admits, $\forall t,j$ an exactly sparse form $||\delta_{t,-j}||_0 \leq s.$ 
    \item \text{Sparsity rate}: The sparsity index obeys $s^2 \log^2 \left(p \lor N\right) / \left(\sqrt{N \log (Tp)}\right) \leq \delta_{N,T}$ and the size of the amelioration set obeys $\hat{s}_3 \leq C\left(1 \lor \hat{s}_1 \lor \hat{s}_2\right).$ Additionally, $\log^3 p/N \leq \delta_{N,T}.$
\end{enumerate}
\end{assumption}

Assumption ASR(i) extends Assumption \ref{assump_dsl_consist_main}(ii) to include sparsity of the DSL first stage. Assumption ASR(ii) controls the rate between sparsity and the asymptotic terms $p,N,T;$ additionally, it constrains the amelioration set to not be substantially larger than the variables selected by the LASSO procedures.

Next, we constrain the minimum and maximum $m$-spare eigenvalues as whenever $p>N$ the empirical design matrix $\mathbb{E}_N [z_{i,t} z_{i,t}^\top ]$ will not have full rank. Define the minimal and maximal $m$-sparse eigenvalue of a semi-definite matrix $M$ as 
\begin{equation*}
    \phi_{\min} (m) [M] \coloneqq \min_{1 \leq || \delta ||_0 \leq m} \frac{\delta^\top M \delta}{|| \delta ||^2} \ \mathrm{and} \ \phi_{\max} (m) [M] \coloneqq \max_{1 \leq || \delta ||_0 \leq m} \frac{\delta^\top M \delta}{|| \delta ||^2}.
\end{equation*}

\begin{assumption}[SE: Sparse Eigenvalues]
\label{SE}
There exists an absolute sequence $l_N \rightarrow \infty$ and such that with probability of at least $1-\Delta_{N,T}$ the maximal and minimal $l_N s $-sparse eigenvalues are bounded from above and away from zero. That is, for absolute constants $0<\kappa' < \kappa'' < \infty,$
$$\kappa' \leq  \phi_{\min} (l_N s) [ \mathbb{E}_N [z_{i,t} z_{i,t}^\top] ] \leq \phi_{\max} (l_N s) [ \mathbb{E}_N [z_{i,t} z_{i,t}^\top] ] \leq \kappa''$$
Similarly, for $\bar{z}_i \coloneqq T^{-1} \sum_i z_{i,t},$ we have
$$\kappa' \leq  \phi_{\min} (l_N s) [ \mathbb{E}_N [\bar{z}_i \bar{z}_i^\top] ] \leq \phi_{\max} (l_N s) [ \mathbb{E}_N [\bar{z}_i \bar{z}_i^\top] ] \leq \kappa''.$$
\end{assumption}

Requiring the minimum $m-$sparse eigenvalue to be bounded away from zero is equivalent to assuming all empirical design submatrices formed by any $m$ components of $z_{i,t}$ are positive definite. 

Next, we impose moment conditions on the structural errors and regressors.

\begin{assumption}[SM: Structural Moments]
\label{SM}
There are absolute constants $0<\omega<\Omega<\infty$ and $4\leq \rho <\infty$ such that for $(y_i, \epsilon_i) \coloneqq (r_{i,t+1}, \epsilon_{i,t+1})$ or $(y_i, \epsilon_i) \coloneqq (z_{i,t,-j}, \epsilon^z_{i,t,j})$ we have $\forall i,t,j$:
\begin{enumerate}
    \item $\mathbb{E}[|z_{i,t,j}|^\rho] \leq \Omega, \omega \leq \mathbb{E}[\epsilon_{i,t+1}^2 | z_{i,t,-j}, \epsilon^z_{i,t,j}] \leq \Omega,$ and $\omega \leq \mathbb{E}[(\epsilon^z_{i,t,j})^2 | z_{i,t,-j}] \leq \Omega;$
    \item $\mathbb{E}[|\epsilon_i|^\rho] + \mathbb{E}[y_i^2] + \max_{1\leq k \leq p} \{\mathbb{E}[z_{i,t,-j,k}^2 y_i^2] + \mathbb{E}[|z_{i,t,-j,k}^3 \epsilon_i^3|] + 1/\mathbb{E}[z^2_{i,t-1,-j,k}] \} \leq \Omega,$
    \item $\max_{1 \leq k \leq p} \{ \mathbb{E} [z_{i,t,-j,k}^2 \epsilon_i^2 ] + \mathbb{E} [z_{i,t,-j,k}^2 y_i^2] \} + \max_{1 \leq i \leq N} ||z_{i,t,-j} ||_\infty^2 \frac{s \log (N \lor p)}{N} \leq \delta_{N,T}$ w.p. $1-\Delta_{N,T}.$
    \item Weak dependence between the first- and second-stage errors: There exists a positive constant $M$ such that $\forall p,T,N:$ 
    $$ \left| \sqrt{\frac{1}{N}}  \sum_{i=1}^N \epsilon^{z}_{i,t,j} \epsilon_{i,t+1} \right| \leq M \log(Tp).$$
    \item Uniformly over $t,j,$ we have $\frac{1}{N} \sum_i (\epsilon^z_{t,j,i} )^2 \xrightarrow[]{p} \mathbb{Z}_{t,j}^0$ for non-stochastic real-valued scalar finite constant $\mathbb{Z}_{t,j}^0,$ which is bounded away from zero.
\end{enumerate}
\end{assumption}

Assumptions (SM)(i)-(iii) are standard for DSL to bound various moments of the errors, characteristics, and returns. Assumption SM(iv) is novel and bounds the dependence between the first- and second-stage errors in the DSL model, which is the non-negligible asymptotic term in the DSL estimation error. This holds trivially for i.i.d. sampling in the cross-section, which we have relaxed to this specified sum. Assumption (v) is novel and introduces a uniform consistency for the second moment of the first-stage errors.

\begin{lemma}
    Under the model \eqref{e:model_dslfm}; Assumption \ref{assump_dsl_consist_main}; and, DSL Assumptions ASR, SE, and SM, the DSL estimator has the property that
    $$ \max_{t,j} | \wh{c}_{t+1,j} - c_{t+1,j} | = O_p\left( \sqrt{\frac{\log (Tp)}{N}} \right).$$
\end{lemma}

\begin{proof}[Proof of Lemma A1]

We proceed with the decomposition of the estimation error using the definition of the DSL estimator and model \eqref{e:model_dslfm}.

\begin{equation*} 
    \begin{aligned}
         \hat{c}_{t+1,j} - c_{t+1,j}  &=  \left( z_{t,j}^\top \mathcal{M}_{\hat{I}} z_{t,j} \right)^{-1} \left( z_{t,j}^\top \mathcal{M}_{\hat{I}} (Z_{t,-j} c_{t+1,-j} + \epsilon_{t+1}) \right) \\
        &=  \left(z_{t,j}^\top \mathcal{M}_{\hat{I}} z_{t,j} \right)^{-1}  \left(Z_{t,-j} \delta_{t,j}\right)^\top \mathcal{M}_{\hat{I}} Z_{t,-j} c_{t+1,-j} \\
        & \qquad +  \left( z_{t,j}^\top \mathcal{M}_{\hat{I}} z_{t,j} \right)^{-1}  \left(Z_{t,-j} \delta_{t,j}\right)^\top \mathcal{M}_{\hat{I}} \epsilon_{t+1} \\
        & \qquad + \left( z_{t,j}^\top \mathcal{M}_{\hat{I}} z_{t,j} \right)^{-1} \epsilon^{z \top}_{t,j} \mathcal{M}_{\hat{I}} Z_{t,-j} c_{t+1,-j} \\
        & \qquad -  \left( z_{t,j}^\top \mathcal{M}_{\hat{I}} z_{t,j} \right)^{-1}  \epsilon^{z \top}_{t,j} \mathcal{P}_{\hat{I}} \epsilon_{t+1} \\
        & \qquad +  \left( z_{t,j}^\top \mathcal{M}_{\hat{I}} z_{t,j} \right)^{-1}    \epsilon^{z \top}_{t,j} \epsilon_{t+1}.
    \end{aligned}
\end{equation*}

From \cite{belloni2014inference} under the aforementioned DSL assumptions, the last term in this five-term decomposition is the asymptotically relevant term while the remaining terms are asymptotically negligible. We first handle the denominator of the fifth term before dealing with the entire term. 
\begin{equation*} 
    \begin{aligned}
        N^{-1} z_{t,j}^\top \mathcal{M}_{\hat{I}} z_{t,j} &= N^{-1} \left(Z_{t,-j} \delta_{t,j} + \epsilon^{z}_{t,j} \right)^\top \mathcal{M}_{\hat{I}} \left( Z_{t,-j} \delta_{t,j} + \epsilon^{z}_{t,j} \right) \\
        &= \epsilon^{z \top}_{t,j} \epsilon^{z }_{t,j} / N + \delta_{t,j}^\top Z_{t,-j}^\top \mathcal{M}_{\hat{I}} Z_{t,-j} \delta_{t,j} / N + 2 \delta_{t,j}^\top Z_{t,-j}^\top \mathcal{M}_{\hat{I}} \epsilon^{z }_{t,j} / N - \epsilon^{z \top}_{t,j}  \mathcal{P}_{\hat{I}}  \epsilon^{z }_{t,j} / N \\
        &\lesssim_P \epsilon^{z \top}_{t,j} \epsilon^{z }_{t,j} / N + o_p(1)
    \end{aligned}
\end{equation*}
where the first equality holds by definition of the first-stage; the second equality holds by multiplying out the terms and by definition of the projection matrices; and, the probabilistic bound holds given the latter three terms are asymptotically negligible, as in the proof of Theorem 1 in \cite{belloni2014inference}, as compared to the sum of second moments of the first-stage errors. Thus, by Assumption SM(v), we conclude  $\epsilon^{z \top}_{t,j} \epsilon^{z }_{t,j} / N$ converges in probability uniformly over $t,j,$ to $\mathbb{Z}_{t,j}^0.$

We proceed with the uniform consistency result.

\begin{equation*} 
    \begin{aligned}
         \max_{t,j} | \wh{c}_{t+1,j} - c_{t+1,j} | & \lesssim_P  \max_{t,j} \left| \left(N^{-1} z_{t,j}^\top \mathcal{M}_{\hat{I}} z_{t,j} \right)^{-1}  \frac{1}{N} \sum_{i=1}^N \epsilon^{z}_{i,t,j} \epsilon_{i,t+1} \right| \\
        & \lesssim_P \sqrt{\frac{1}{N}} \max_{t,j} \left| \sqrt{\frac{1}{N}} \sum_{i=1}^N \epsilon^{z}_{i,t,j} \epsilon_{i,t+1} \right| \\
        & \lesssim_P \sqrt{\frac{\log (Tp)}{N}}
    \end{aligned}
\end{equation*}
which holds for the first probabilistic bound by substituting the decomposition above; for the second probabilistic bound, using the above result to replace the denominator with a constant that is bounded away from zero uniformly over $t,j$; and, the final bound holds by assumption SM(iv). In the case of i.i.d. sampling in the cross-section or if the dependence is sufficiently weak such that SM(iv) holds, then we can invoke Lemma A.4 in \cite{belloni2018high} to conclude the mean zero scalar random variable $\epsilon^{z}_{i,t,j} \epsilon_{i,t+1}$, given the moment conditions of the DSL model, has a maximal deviation that converges in probability to zero at the specified rate if we further constrain the moment of the mean zero random variable $\E \left[ \max_{t,j} \left| \epsilon^{z}_{i,t,j} \epsilon_{i,t+1} \right|^q \right] \leq M^q$ for $q>2$ and absolute constant $M$ uniformly across $t,j.$
\end{proof}
    
\subsection{Consistency of Latent Factors and Loadings}

We first prove a bound on the distance between the infeasible and feasible symmetric matrix used in the eigendecompositions. Let $\wh{\L}_{Tp} \in \R^{k \times k}$ be a diagonal matrix containing the $k$ largest eigenvalues of $(Tp)^{-1} \wh{C} \wh{C}^\top$ and similarly for $\L_{Tp} \in \R^{k \times k},$ a diagonal matrix containing the $k$ largest eigenvalues of $(Tp)^{-1} \wh{C} \wh{C}^\top.$ 

\begin{lemma} Under the assumptions of Lemma A1, $\norm{(Tp)^{-1} \wh{C}\wh{C}^\top - (Tp)^{-1} CC^\top} = O_p \left( \frac{\log Tp}{N} \right).$
\end{lemma}

\begin{proof}[Proof of Lemma A2]
    \begin{equation*} 
        \begin{aligned}
        \norm{\wh{C}\wh{C}^\top - CC^\top} &=  \norm{\wh{C}\wh{C}^\top - C\wh{C}^\top + C\wh{C}^\top - CC^\top} \\
        &\leq   \norm{C\wh{C}^\top - CC^\top} + \norm{\wh{C}\wh{C}^\top - C\wh{C}^\top} \\
        &\leq  \norm{C} \norm{\wh{C} - C} + \norm{\wh{C} - C} \norm{\wh{C}}  \\
        &=  \norm{C} \norm{\wh{C} - C} + \norm{\wh{C} - C} \norm{\wh{C} - C + C}  \\
        &\leq \norm{C} \norm{\wh{C} - C} + \norm{\wh{C} - C} \left(\norm{\wh{C} - C} + \norm{C} \right)  \\
        &= 2 \norm{C} \norm{\wh{C} - C} + \norm{\wh{C} - C}^2 \\
        &\lesssim_P \sqrt{sT}\norm{\wh{C} - C} + \norm{\wh{C} - C}^2 \\
        &\leq \sqrt{spT^2} \max_{t,j} | \wh{c}_{t+1,j} - c_{t+1,j} | + Tp \max_{t,j} | \wh{c}_{t+1,j} - c_{t+1,j} |^2 \\
        &\lesssim_P \sqrt{\frac{spT^2 \log (Tp)}{N}} + \frac{Tp \log (Tp)}{N} \lesssim \frac{Tp \log (Tp)}{N}.
        \end{aligned}
    \end{equation*}
    where the first and third inequality holds by the triangle inequality; the second inequality holds by Cauchy-Schwarz; the first probabilistic bound holds by Assumption \ref{assump_dsl_consist_main}(i) / \ref{ASR}(i); and, the last probabilistic bound holds by Lemma A1. We use the final bound for simplicity.
\end{proof}

We next bound the estimation error between the feasible and infeasible eigenvalues.

\begin{lemma} Under the assumptions of Lemma A2 and Assumption \ref{assump_number_factors}, $$\norm{\wh{\Lambda}_{Tp} - \L_{Tp}}^2 = O_p\left(\frac{s^2 \log^2 (Tp)}{N^2}\right).$$
\end{lemma}

\begin{proof}[Proof of Lemma A3] 
    \begin{equation*} 
        \begin{aligned}
        \norm{\wh{\Lambda}_{Tp}- \L_{Tp}}^2 &= \sum_{l=1}^k \sum_{i=1}^k (\wh{\l}_l - \l_i)^2 \\
        &\leq k^2 \max_{l \in \{1,\dots,k\}} | \wh{\l}_l - \l_l |^2 \\ 
        &\leq k^2 \max_{l \in \{1,\dots,T\}} | \wh{\l}_l - \l_l |^2 \\
        &\leq \frac{k^2}{T^2 p^2} \norm{\wh{C}\wh{C}^\top - C C^\top}^2 = O_p(\frac{s^2 \log^2 (Tp)}{N^2})
        \end{aligned}
    \end{equation*}
    where the first equality is the definition of the Frobenius norm; the first inequality bounds the sum by the maximum element; the second inequality bounds the maximum deviation between the $k$ largest eigenvalues of the feasible and infeasible decompositions by the deviations between all $T$ eigenvalues; the last inequality controls the stability of the spectrum by applying Weyl's inequality from Theorem 4.5.3 of \cite{vershynin2018high}; and, the probabilistic bound holds by Lemma A2.
\end{proof}

We next prove a lemma for the time series average of the $\ell_2$ norm of the feasible and infeasible eigenvectors.

\begin{lemma}
    Under Assumption \ref{assump_latent_fm}(iii) and those of Lemma A3, there exists an orthogonal matrix $\wh{O}\in \R^{k \times k}$ such that $\norm{\wh{F} - \wt{F} \wh{O}^\top}^2 = O_p\left(\frac{\log^2 (Tp)}{N^2}\right).$
\end{lemma}
    
\begin{proof}[Proof of Lemma A4]
    We use a variant of the Davis-Kahan theorem shown in \cite{yu2015useful} where
    $$\delta \coloneqq \min_{i:i\neq l} |\lambda_l- \lambda_i | > 0,$$
    which holds by Assumption \ref{assump_latent_fm}(iii), to conclude for some $\wh{O} \in \R^{k \times k}$ orthogonal matrix that
    \begin{equation*} 
        \begin{aligned}
        2^{3/2} \delta^{-1} (Tp)^{-2} \norm{\wh{C}\wh{C}^\top - CC^\top}^2 &\geq || \wh{F} \wh{O} - \wt{F} ||^2 
        = || \wh{F}  - \wt{F} \wh{O}^\top ||^2
        \end{aligned}
    \end{equation*}
    where the inequality is the use of the variant of the Davis-Kahan theorem, bounding the distance between the eigenvectors by the distance between the original matrices, and the equality follows given post multiplying by an orthogonal matrix does not change the Frobenius norm. The rate in the result then follows given Lemma A2.
\end{proof}

We next bound the $\ell_2$ norm between the feasible and infeasible eigenvectors.

\begin{lemma}
    Under the assumptions of Lemma A4, $\norm{\wh{f}_{t+1} - \wh{O}^\top \wt{f}_{t+1}} = O_p\left(\sqrt{\frac{s \log(Tp)}{N}}\right).$
\end{lemma}

\begin{proof}[Proof of Lemma A5] First, we perform the following decomposition using the definition of the eigenvectors.
    \begin{equation*}
        \begin{aligned}
            \wh{f}_{t+1} - \wh{O}^\top \wt{f}_{t+1} &= (Tp)^{-1} \wh{\L}_{Tp}^{-1} \wh{F}^\top \wh{C} \wh{C}_{t+1} -  (Tp)^{-1} \wh{O}^\top \L_{Tp}^{-1} \wt{F}^\top C C_{t+1} \\
            &= (Tp)^{-1} \left( \wh{\L}_{Tp}^{-1} - \L_{Tp}^{-1} \right) \wh{F}^\top \left( \wh{C} \wh{C}_{t+1} - C C_{t+1} \right) \\
            & \qquad + (Tp)^{-1} \wh{O}^\top \L_{Tp}^{-1} \left( \wt{F}^\top - \wh{F}^\top \right) C C_{t+1} \\
            & \qquad + (Tp)^{-1}  \left( \wh{\L}_{Tp}^{-1} - \wh{O}^\top \L_{Tp}^{-1} \right) \wh{F}^\top C C_{t+1} \\
            & \qquad + (Tp)^{-1}  \L_{Tp}^{-1} \wh{F}^\top \left( \wh{C} \wh{C}_{t+1} - C C_{t+1} \right)
        \end{aligned}
    \end{equation*}
    where the equality follows by adding and subtracting terms.

    Thus,
    \begin{equation*}
        \begin{aligned}
            \norm{\wh{f}_{t+1} - \wh{O}^\top \wt{f}_{t+1} } &\leq (Tp)^{-1} \norm{ \wh{\L}_{Tp}^{-1} - \L_{Tp}^{-1}} \norm{ \wh{F} } \norm{ \wh{C} \wh{C}_{t+1} - C C_{t+1} } \\
            & \qquad + (Tp)^{-1} \norm{ \L_{Tp}^{-1} } \norm{\wh{F} - \wt{F} } \norm{  C C_{t+1} } \\
            & \qquad + (Tp)^{-1}  \norm{ \wh{\L}_{Tp}^{-1} - \L_{Tp}^{-1} } \norm{ \wh{F} } \norm{ C C_{t+1} } \\
            & \qquad + (Tp)^{-1} \norm{ \L_{Tp}^{-1} }  \norm{ \wh{F} }  \norm{\wh{C} \wh{C}_{t+1} - C C_{t+1} } \\
            &=  O_p\left(\sqrt{\frac{s^3  \log(Tp)^3}{p N^{3}}}\right) + 
            O_p\left(\frac{\sqrt{T} s \log (Tp)}{Tp N}\right) \\
            & \qquad + O_p\left(\frac{s^2  \log(Tp)}{p N}\right) + O_p\left(\sqrt{\frac{s\log(Tp) }{N}}\right)  \\
            &= O_p\left(\sqrt{\frac{s \log(Tp)}{N}}\right)
        \end{aligned}
    \end{equation*}
    where the first inequality follows from the aforementioned decomposition in this proof with the use of the triangle and Cauchy-Schwarz inequalities; the first probabilistic bound holds given the normalization that $\wh{F}^\top \wh{F}/T = I_{k}$ then $|| \wh{F} || = \sqrt{Tk},$ given $\L_{Tp}$ contains $k$ nonzero real-valued eigenvalues bounded away from zero by Assumption \ref{assump_latent_fm}(iii), given the rates from Lemmas A3 and A4 (which gives same rate by CMT), given---similar to lemma A2---$|| \wh{C} \wh{C}_{t+1} - C C_{t+1}|| = O_p(\sqrt{\frac{sTp^2 \log(Tp)}{N}}),$ and $||CC_{t+1}|| = O_p(\sqrt{s^2 T})$ by Assumption \ref{assump_dsl_consist_main}(ii); and, the final probabilistic bound holds for simplicity of exposition.
\end{proof}

\begin{lemma}
    Under the assumptions of A5, for $H^\top = \wh{O}^\top \L_{Tp}^{-1} (F^\top F^0 /T) (\G_\b^{0 \top} \G_\b^0 / p)$ we have $$\wh{O} \wt{f}_{t+1} - H^\top f^0_{t+1} = 0.$$
\end{lemma}

\begin{proof}[Proof of Lemma A6] 
\begin{equation*} 
    \begin{aligned}
    \wh{O} \wt{f}_{t+1} &= (Tp)^{-1} \wh{O}^\top \L_{Tp}^{-1} F^\top C C_{t+1} \\
    &= (Tp)^{-1} \wh{O}^\top \L_{Tp}^{-1} F^\top F \G_\b^\top \G_\b f_{t+1} \\
    &= (Tp)^{-1} \wh{O}^\top \L_{Tp}^{-1} F^\top F^0 \G_\b^{0 \top} \G^{0 \top}_\b f^0_{t+1} \\
    &= H^\top f^0_{t+1}.
    \end{aligned}
\end{equation*}
where the first equality holds by the definition of the infeasible eigenvectors; the second equality holds given the definition of $C;$ the third equality holds given the definitions of the true loadings and factors as rotations of the observed ones; and, the final equality holds by definition of the $H$ matrix.
\end{proof}

Finally, using the above lemmas, we prove Propositions 1 and 2.
    
\begin{proof}[Proof of Proposition 1] 
\begin{equation*} 
    \begin{aligned}
    \norm{\wh{f}_{t+1} - H^\top f_{t+1}^0} &\leq \norm{\wh{f}_{t+1} - \wh{O}^\top \wt{f}_{t+1}} + \norm{\wh{O}^\top \wt{f}_{t+1} - H^\top f_{t+1}^0} \\
    &= O_p\left(\sqrt{\frac{s\log(Tp)}{N}}\right)
    \end{aligned}
\end{equation*}
by Lemmas A5 and A6.
\end{proof}

Next, we provide a lemma for the $\ell_\infty$ norm of the PCA estimation error for the loadings, which will allow us to obtain norms on the soft-threshold estimation error for the loadings by use of a tool from the high-dimensional econometrics handbook \citep{belloni2018high}.

\begin{lemma}
    Under the assumptions of Lemma A1 and Assumption \ref{assump_number_factors}, $$\norm{\wh{\G}_\b - \G_\b^0 (H^\top )^{-1} }_\infty = O_p \left( \sqrt{ \frac{\log(Tp)}{N} } \right).$$
\end{lemma}

\begin{proof}[Proof of Lemma A7]
\begin{equation*}
    \begin{aligned}
        \norm{\wh{\G}_\b - \G_\b^0 (H^\top )^{-1} }_\infty &= \max_j \left( \sum_{l=1}^k \left| \wh{\G}_{\b,j,l} - (\G_{\b,j}^0)^\top (H^\top)^{-1}_l \right| \right) \\
        &= \max_j \norm{\wh{\G}_{\b,j} - H^{-1} \G^0_{\b,j} }_1 \\
        &= \max_j \norm{\wh{\G}_{\b,j} \pm \wt{\G}_{\b,j} - H^{-1} \G^0_{\b,j} }_1 \\
        &\leq \max_j \norm{\wh{\G}_{\b,j} - \wt{\G}_{\b,j} }_1 + \max_j \norm{\wt{\G}_{\b,j} - H^{-1} \G^0_{\b,j} }_1 \\
        &= \max_j \norm{T^{-1} \wh{F}^\top \wh{C}_j - T^{-1} F^\top C_j }_1 + \max_j \norm{T^{-1} F^\top C_j - H^{-1} \G^0_{\b,j} }_1 \\
        &= T^{-1} \max_j \norm{\wh{F}^\top \wh{C}_j \pm \wh{F}^\top C_j - F^\top C_j }_1 + \max_j \norm{T^{-1} F^\top F \G_{\b,j} - H^{-1} \G^0_{\b,j} }_1
    \end{aligned}
\end{equation*}

\begin{equation*}
    \begin{aligned}
        &\leq T^{-1} \max_j \norm{\wh{F}^\top \left( \wh{C}_j - C_j \right) }_1 + T^{-1} \max_j \norm{\left(\wh{F} - F\right)^\top C_j }_1 \\
        & \qquad + \max_j \norm{T^{-1} F^\top F \G_{\b,j} - H^{-1} \G^0_{\b,j} }_1 \\
        &\leq \frac{\sqrt{k}}{T} \norm{\wh{F}}_2 \max_j \norm{\wh{C}_j - C_j }_2 + \frac{\sqrt{k}}{T}  \norm{\wh{F} - F }_2 \max_j \norm{ C_j }_2 \\
        & \qquad + \sqrt{k} \norm{T^{-1} F^\top F - I_{k \times k} }_2 \norm{H^{-1}}_2 \max_j \norm{ \G^0_{\b,j} }_2 \\
        &\lesssim_P \max_{t,j} \left| \wh{c}_{t+1,j} - c_{t+1,j} \right| + T^{-1/2} \norm{\wh{F} - F } \\
        &\lesssim_P O_p \left( \sqrt{\frac{\log(Tp)}{N}} \right) + O_p \left( \frac{\log(Tp)}{N} \right) \\
    \end{aligned}
\end{equation*}
where the first two equalities follow from the definition of the $\ell_\infty$ norm; the third equality adds and subtracts; the first inequality uses the triangle inequality; the fourth equality uses the definitions of the feasible and infeasible estimator; the second inequality uses a triangle inequality; the last inequality uses $\norm{A^\top x}_1 \leq \sqrt{k} \norm{A}_2 \norm{x}_2$; the first probabilistic bound uses $\norm{\wh{F}}_2 = O_p (\sqrt{kT}),$ the sum of the estimation errors in the characteristic portfolios, $|\wh{c}_{t+1,j} - c_{t+1,j}|$ is bounded by $T$ times the maximum element, the elements of $C_j$ are bounded random variables hence $\max_j \norm{C_j} = O_p(\sqrt{T}),$ $T^{-1} \sum_t f^0_{t+1} f_{t+1}^{0,\top} \rightarrow_p \Sigma_f$ by Assumption \ref{assump_latent_fm}(i), $\norm{H^{-1}} = O_p(1)$ by definition of the invertible $H$ matrix, and $\G_\b^0$ contains bounded elements uniformly over $j$ by Assumption \ref{assump_latent_fm}(ii); and, the last probabilistic bound follows by Lemmas A1 and A4 and assumption \ref{ASR}(ii).
\end{proof}

\begin{proof}[Proof of Proposition 2]
In Lemma A7, we show $\norm{\wh{\G}_\b - \G_\b^0 (H^\top )^{-1} }_\infty \lesssim_P \sqrt{\frac{\log(Tp)}{N}},$ which allows us to invoke Theorem 2.10 from \cite{belloni2018high} under exact sparsity of $\G_\b^0$ where $\lambda,$ the hyperparameter used to soft-threshold the $\ell_1$ norm of the rows of $\wh{\G}_\b,$ is selected such that with probability approaching 1, $$\lambda \geq (1-\alpha) - \text{quantile} \ \text{of} \ \norm{\wh{\G}_\b - \G_\b^0 (H^\top )^{-1} }_\infty.$$ That is, $\lambda$ can be set to the product of a large constant and the rate of the $\ell_\infty$ norm, $\sqrt{\frac{\log(Tp)}{N}},$ to ensure this inequality holds.\footnote{In practice, we cross-validate for a finite-sample optimal $\lambda.$} Then, by Theorem 2.10 given $\alpha \rightarrow 0$ and $\lambda \lesssim \sqrt{\log(Tp)/N},$ we have for all $q \geq 1$
\begin{equation*} 
    \begin{aligned}
        \norm{ \check{\G}_{\b,l} - \G_{\b}^0 (H^\top )^{-1}_l }_q \lesssim_P s^{1/q} \sqrt{\frac{\log(Tp)}{N}}.
    \end{aligned}
\end{equation*}
This holds column-by-column for the matrix estimation error $\check{\G}_{\b} - \G_\b^0,$ which we can thus square and sum together for the squared Frobenius norm of the estimation error at the same rate.
\end{proof}

\subsection{Consistency and Normality of Observable Factor Risk Premia}

In these results, we are using the eigenvectors and loadings derived from the demeaned characteristic portfolio matrix: $\wh{C}^D \coloneqq \wh{C} - \iota_T T^{-1} \sum_t \wh{c}_{t+1}^\top.$ That is, $\wh{V}$ are the $\sqrt{T}$ scaled eigenvectors associated with the $k$ largest eigenvalues of $(Tp)^{-1} \wh{C}^D \wh{C}^{D\top}.$ Further, $\wh{\G}_\beta^D = T^{-1} \wh{C}^{D\top} \wh{V}.$ The results established in the above subsection would follow analogously for this new notation as we simply have mean zero eigenvectors. Finally, our cross sectional and time series OLS estimators are standard:
\begin{equation*} 
    \wh{\g} \coloneqq \left( \bar{\wh{\b}}^\top \bar{\wh{\b}} \right)^{-1} \bar{\wh{\b}}^\top \bar{r}, \qquad
    \wh{\eta} \coloneqq \left( \wh{V}^\top \wh{V} \right)^{-1} \wh{V}^\top G,
\end{equation*}
where $\bar{\wh{\b}} \coloneqq \bar{Z} \check{\G}_\b^D$ for $\bar{Z} = T^{-1} \sum_t Z_{t}$ for $Z_{t} \in \mathbb{R}^{N \times p}, \ \forall t.$ The same holds for the time series average return $\bar{r} \in \mathbb{R}^N.$

\begin{lemma}
    Under the models \eqref{e:model_dslfm} and \eqref{e:model_ob_risk_premia}; Assumptions \ref{assump_dsl_consist_main}, \ref{assump_latent_fm}, and \ref{assump_number_factors}; and, DSL Assumptions ASR, SE, and SM; we have $$\norm{ N^{-1} \bar{\wh{\b}}^\top \bar{\wh{\b}} - N^{-1} \bar{\b}^\top \bar{\b}} = O_p\left(\sqrt{\frac{s^2 \log(Tp)}{N} } \right).$$
\end{lemma}

\begin{proof}[Proof of Lemma A8]
\begin{equation*} 
    \begin{aligned}
        \norm{ N^{-1} \bar{\wh{\b}}^\top \bar{\wh{\b}} - N^{-1} \bar{\b}^\top \bar{\b}} &= N^{-1} \norm{  \bar{\wh{\b}}^\top \bar{\wh{\b}} \pm \bar{\wh{\b}}^\top \bar{\b} - \bar{\b}^\top \bar{\b} }  \\
        &\leq N^{-1} \norm{  \bar{\wh{\b}}^\top \bar{\wh{\b}} - \bar{\wh{\b}}^\top \bar{\b} } + N^{-1} \norm{\bar{\wh{\b}}^\top \bar{\b} - \bar{\b}^\top \bar{\b} } \\
        &= N^{-1} \norm{  \left( \bar{\wh{\b}} \pm  \bar{\b} \right)^\top \left( \bar{\wh{\b}} - \bar{\b} \right) } + N^{-1} \norm{ \left( \bar{\wh{\b}}  - \bar{\b} \right)^\top \bar{\b} } \\
    \end{aligned}
\end{equation*}

\begin{equation*} 
    \begin{aligned}
        &\leq N^{-1} \norm{  \bar{\wh{\b}} - \bar{\b} }^2 + \frac{2}{N} \norm{ \left( \bar{\wh{\b}}  - \bar{\b} \right)^\top \bar{\b} }  \\
        &= N^{-1} \norm{  \bar{Z} \left(  \check{\G}_\b^D -  \G_\b^0 \right) }^2 + \frac{2}{N} \norm{   \left(  \check{\G}_\b^D -  \G_\b^0 \right)^\top \bar{Z}^\top \bar{Z} \G_\b^0 } \\
        &\leq \norm{ \left(  \check{\G}_\b^D -  \G_\b^0 \right)^\top \frac{\bar{Z}^\top \bar{Z}}{N} \left(  \check{\G}_\b^D -  \G_\b^0 \right) }^2 \\
        & \qquad +  2 \norm{   \left(  \check{\G}_\b^D -  \G_\b^0 \right)^\top \frac{\bar{Z}^\top \bar{Z}}{N} \G_\b^0 } \\
        &\leq \norm{ \check{\G}_{\b,l}^D -  \G_{\b,l}^0 }^2 \norm{\check{\G}_\b^D -  \G_\b^0}^2 \phi^2_{\max} (2s) \left[ \frac{\bar{Z}^\top \bar{Z}}{N} \right] \\
        & \qquad + 2 \norm{ \check{\G}_{\b,l}^D -  \G_{\b,l}^0 } \norm{\G_\b^0} \phi_{\max} (2s) \left[ \frac{\bar{Z}^\top \bar{Z}}{N} \right] \\
        &\lesssim_P O_p \left( \frac{s^2 \log^2 (Tp) }{N^2} \right) + O_p \left( \sqrt{\frac{s^2 \log(Tp)}{N}} \right)
    \end{aligned}
\end{equation*}
where the first equality follows by adding and subtracting; the first inequality follows by the triangle inequality; the second equality follows from adding and subtracting and rearranging; the second inequality follows from the triangle and Cauchy-Schwartz inequalities; the third equality follows from the definition of the factor loading estimator and the average factor loading; the third inequality follows from $||A|| \leq ||AA\top ||;$ the final inequality follows from multiplying and dividing by the norms to obtain unit vectors and then bounding with the $s+\hat{s}$-maximally sparse eigenvalue of $\frac{\bar{Z}\top \bar{Z}}{N}$ (where the $s+\hat{s} \leq 2s$ by thresholding estimator); and, the probabilistic bound holds given the maximum eigenvalue is bounded by DSL Assumption SE, $\norm{\G_\b^0} = O_p(\sqrt{s})$ by Assumption \ref{assump_dsl_consist_main}(ii), and $\norm{  \check{\G}_{\b,l}^D -  \G_{\b,l}^0 } = O_p\left( \sqrt{\frac{s \log(Tp)}{N}}\right),$ which is the same rate for the entire matrix per Proposition 2.
\end{proof}

\begin{lemma}
    Under the assumptions of Lemma A8, we have $$\frac{\sqrt{T}}{N} \norm{\bar{\wh{\b}}^\top \bar{r} - \bar{\b}^\top \bar{r} } =  O_p \left( \sqrt{\frac{T s^2 \log (Tp)}{N}} \right).$$
\end{lemma}

\begin{proof}[Proof of Lemma A9]
\begin{equation*} 
    \begin{aligned}
        \frac{\sqrt{T}}{N} \norm{\bar{\wh{\b}}^\top \bar{r} - \bar{\b}^\top \bar{r}} &= \frac{\sqrt{T}}{N} \norm{\left(\check{\G}_\b^D -  \G_\b^0 \right)^\top \bar{Z}^\top \left( \bar{Z} \G_\b^0 \g^0 + T^{-1} \sum_t Z_t \G_\b^0 v_{t+1}^0 + \bar{\e} \right) } \\
        &\leq \frac{\sqrt{T}}{N} \norm{\left(\check{\G}_\b^D -  \G_\b^0 \right)^\top \bar{Z}^\top  \bar{Z} \G_\b^0 \g^0 } \\
        & \qquad + \frac{\sqrt{T}}{N} \norm{\left(\check{\G}_\b^D -  \G_\b^0 \right)^\top \bar{Z}^\top T^{-1} \sum_t Z_t \G_\b^0 v_{t+1}^0 } \\
        & \qquad + \frac{\sqrt{T}}{N} \norm{\left(\check{\G}_\b^D -  \G_\b^0 \right)^\top \bar{Z}^\top \bar{\e}  } \\
        &\leq \sqrt{T} \norm{\check{\G}_{\b,l}^D -  \G_{\b,l}^0} \norm{\G_\b^0 \g^0} \phi_{\max} (2s) \left[ \frac{\bar{Z}^\top \bar{Z}}{N} \right] \\
        & \qquad + \norm{ \check{\G}_{\b,l}^D -  \G_{\b,l}^0 }^2  \phi_{\max} (2s) \left[ \frac{\bar{Z}^\top \bar{Z}}{N} \right] \sqrt{\frac{T}{N}} \norm{T^{-1} \sum_t Z_t \G_\b^0 v_{t+1}^0 } \\
        & \qquad + \norm{ \check{\G}_{\b,l}^D -  \G_{\b,l}^0 }^2 \phi_{\max} (2s) \left[ \frac{\bar{Z}^\top \bar{Z}}{N} \right] \sqrt{\frac{T}{N}} \norm{\bar{\e}} \\
        & \lesssim_p O_p \left( \sqrt{\frac{T s^2 \log (Tp)}{N}} \right) + O_p \left( \frac{s^{3/2} \log (Tp)}{N} \right) \\
        & \qquad + O_p \left( \frac{s \log (Tp)}{N} \right)
    \end{aligned}
\end{equation*}
where the first equality follows by the definitions of the factor loading estimator, the average factor loading, and the time series average return of assets; the first inequality follows by the triangle inequality; the second inequality holds by an analogous argument to Lemma A8; and, the probabilistic bound holds again by an analogous argument to Lemma A8 with the additional bounds of $\norm{\G_\b^0 \g^0} = O_p(\sqrt{s})$ by Assumptions \ref{assump_latent_fm}(i)-(ii) and \ref{assump_dsl_consist_main}(ii), $\norm{T^{-1} \sum_t Z_t \G_\b^0 v_{t+1}^0 } = O_p(\sqrt{\frac{sN}{T}})$ by Assumption \ref{assump_inf}(ii), and $\norm{\bar{\e}} = O_p(\sqrt{\frac{N}{T}})$ by Assumption \ref{assump_inf}(i).
\end{proof}

\begin{lemma}
    Under the assumptions of Lemma A8 and Assumption \ref{assump_inf}(iii), we have
    $$ \sqrt{T} \left( \wh{\g} - H \g_0 \right) = O_p(1).$$
\end{lemma}

\begin{proof} [Proof of Lemma A10]
We decompose the estimation error into three terms.
\begin{equation*}
    \begin{aligned}
        \sqrt{T} \left( \wh{\g} - H \g_0 \right) &= \sqrt{T} \left( \wh{\g} - \wt{\g} \right) + \sqrt{T} \left( \wt{\g} - H \g_0 \right) \\
        &= \left( \frac{\bar{\wh{\b}}^\top \bar{\wh{\b}}}{N} \right)^{-1} \frac{\sqrt{T} \bar{\wh{\b}}^\top \bar{r}}{N} \pm \left( \frac{\bar{\b}^\top \bar{\b}}{N} \right)^{-1} \frac{\sqrt{T} \bar{\b}^\top \bar{r}}{N} - \sqrt{T} H \g_0 \\
        &= \underbrace{\left( \frac{\bar{\wh{\b}}^\top \bar{\wh{\b}}}{N} \right)^{-1} \frac{\sqrt{T} \bar{\wh{\b}}^\top \bar{r}}{N} - \left( \frac{\bar{\b}^\top \bar{\b}}{N} \right)^{-1} \frac{\sqrt{T} \bar{\b}^\top \bar{r}}{N}}_{\mathcal{A}_\g} \\
        & \qquad + \underbrace{\sqrt{T} H^\top \left(  \frac{\G_\b^{0\top} \bar{Z}^\top \bar{Z} \G_\b^0}{N} \right)^{-1} \frac{ \G_\b^{0\top} \bar{Z}^\top}{N} \frac{1}{T} \sum_t Z_t \G_\b^0 v_{t+1}^0}_{\mathcal{B}_\g} \\
        & \qquad + \underbrace{\sqrt{T} H^\top \left(  \frac{\G_\b^{0\top} \bar{Z}^\top \bar{Z} \G_\b^0}{N} \right)^{-1} \frac{ \G_\b^{0\top} \bar{Z}^\top}{N} \bar{\e}}_{\mathcal{C}_\g} \\
        &=: \mathcal{A}_\g + \mathcal{B}_\g + \mathcal{C}_\g.
    \end{aligned}
\end{equation*}
where the first equality follows by adding and subtracting; the second equality follows from the definition of the feasible and infeasible estimator; the last equality follows from the definition of the time series average of asset returns and rearranging; and, finally, we define three terms to prove, in the rest of this proof, $\mathcal{A}_\g + \mathcal{C}_\g = o_p(1)$ and $\mathcal{B}_\g = O_p(1)$ to prove the lemma.

First, we prove $\mathcal{A}_\g = o_P(1).$ Define notation:
$$\mathcal{A}_\g \coloneqq \wh{\mathcal{A}}^{-1}_i \wh{\mathcal{A}}_{ii} - \mathcal{A}_i^{-1} \mathcal{A}_{ii} \coloneqq \left( \frac{\bar{\wh{\b}}^\top \bar{\wh{\b}}}{N} \right)^{-1} \frac{\sqrt{T} \bar{\wh{\b}}^\top \bar{r}}{N} - \left( \frac{\bar{\b}^\top \bar{\b}}{N} \right)^{-1} \frac{\sqrt{T} \bar{\b}^\top \bar{r}}{N}$$
and $\Delta_i \coloneqq \wh{\mathcal{A}}^{-1}_i - \mathcal{A}_i^{-1}$ and $\Delta_{ii} \coloneqq \wh{\mathcal{A}}_{ii} - \mathcal{A}_{ii}.$ Thus,
\begin{equation*} 
    \begin{aligned}
        \wh{\mathcal{A}}^{-1}_i \wh{\mathcal{A}}_{ii} - \mathcal{A}_i^{-1} \mathcal{A}_{ii} = \mathcal{A}_{ii} \Delta_i + \mathcal{A}^{-1}_i \Delta_{ii} + \Delta_i \Delta_{ii} = O_p(1) o_p(1) + O_p(1) o_p(1) + o_p(1) o_p(1) = o_p(1)
    \end{aligned}
\end{equation*}
given: by Lemma A9, $\mathcal{A}_{ii}=O_p(1)$ and $\Delta_{ii} = o_p(1);$ and, by Lemma A8 and CMT, $\mathcal{A}_i = O_p(1)$ and $\Delta_{i}=o_p(1).$

Second, $\mathcal{B}_\g=O_p(1)$ given
\begin{equation*} 
    \begin{aligned}
        \norm{\mathcal{B}_\g } &= \sqrt{T} \norm{ H^\top \left(  \frac{\G_\b^{0\top} \bar{Z}^\top \bar{Z} \G_\b^0}{N} \right)^{-1} \frac{ \G_\b^{0\top} \bar{Z}^\top}{N} \frac{1}{T} \sum_t Z_t \G_\b^0 v_{t+1}^0 } \\
        & \leq \norm{H} \norm{\left(  \frac{\G_\b^{0\top} \bar{Z}^\top \bar{Z} \G_\b^0}{N} \right)^{-1}}_2 \norm{\frac{\G_\b^{0\top} \bar{Z}^\top}{\sqrt{N}} } \norm{\sqrt{\frac{1}{NT}} \sum_t Z_t \G_\b^0 v_{t+1}^0} \\
        & \lesssim_P \phi^{-1}_{\min} (2s) [ N^{-1} \bar{Z}^\top \bar{Z} ] 
    \end{aligned}
\end{equation*}
where the first equality substitutes the notation; the first inequality follows by $\norm{ A B x }\leq \norm{A} \norm{B}_2 \norm{x}$ for matrices $A,B$ and vector $x$ where $\norm{B}_2$ is the spectral norm; and, the probabilistic bound follows given $\norm{H}=O_p(1),$ the spectral norm of the inverse matrix is bounded by the $2s$-sparse minimum eigenvalue of $N^{-1} \bar{Z}^\top \bar{Z}$ which is bounded away from zero by DSL Assumption SE, $\norm{\bar{Z} \G_\b^0 } = O_p(1)$ by an analogues argument using the $2s$-sparse maximum eigenvalue, and $\norm{ \sqrt{\frac{1}{TN} } \sum_t Z_t \G_\b^0 v_{t+1}^0 } = O_p\left( 1 \right)$ by Assumption \ref{assump_inf}(ii).

Third, $\mathcal{C}_\g=o_p(1)$ given
\begin{equation*} 
    \begin{aligned}
        \norm{\mathcal{C}_\g } &= \norm{ \sqrt{T} H^\top \left(  \frac{\G_\b^{0\top} \bar{Z}^\top \bar{Z} \G_\b^0}{N} \right)^{-1} \frac{ \G_\b^{0\top} \bar{Z}^\top}{N} \bar{\e} } \\
        & \lesssim_P \norm{ \frac{ \G_\b^{0\top} \left( \bar{Z} \pm \E [ \bar{Z} ] \right)^\top}{N} \sqrt{\frac{1}{T}} \sum_t \epsilon_t }  \\
        &= \norm{ \sqrt{\frac{1}{T}} \sum_t \sum_{j=1}^s \frac{1}{N} \sum_i \G_{\b,j}^0 \left( \bar{Z}_{i,j} - \E [ \bar{Z}_{i,j} ] \right)^\top  \epsilon_{i,t+1} } \\
        & \qquad + \norm{  \sqrt{\frac{1}{T}} \sum_t \sum_{j=1}^s \frac{1}{N} \sum_i \G_{\b,j}^0 \E [ \bar{Z}_{i,j} ]^\top  \epsilon_{i,t+1} }  \\
        &= O_p(\sqrt{\frac{T}{N}}) = o_p(1)
    \end{aligned}
\end{equation*}
where the first equality substitutes the notation; the first probabilistic bound holds by adding and subtracting $\E[\bar{Z}]$ and the previously established results on $\norm{H}$ and the spectral norm of the inverse design matrix; the second equality holds by opening up the matrix multiplication for the two terms noting $\G_\b^0$ selects only $s$ rows of $\bar{Z}^\top;$ the final probabilistic bound holds given Assumption \ref{assump_inf}(i) yields $N^{-1} \sum_i \e_{i,t+1} = O_p(N^{-1/2})$ and a LLN on $T^{-1} \sum_t \bar{Z}_{i,j} - T^{-1} \sum_t \E [ \bar{Z}_{i,j} ] = o_p(1)$ using the structural moment Assumptions \ref{SM}; and, the final assumption holds given $\sqrt{T / N} \rightarrow 0.$

Thus, given $\mathcal{A}_\g + \mathcal{B}_\g + \mathcal{C}_\g = o_p(1) + O_p(1) + o_p(1) = O_p(1),$ the lemma holds.
\end{proof}

\begin{lemma}
    Under the assumptions of Lemma A8, we have $$\norm{T^{-1} \hat{V}^\top \hat{V} - T^{-1} V^\top V} = O_p\left(\sqrt{\frac{ \log^2(Tp)}{T N^2} } \right).$$
\end{lemma}

\begin{proof}[Proof of Lemma A11]
\begin{equation*} 
    \begin{aligned}
        \norm{T^{-1} \hat{V}^\top \hat{V} - T^{-1} V^\top V} &= T^{-1} \norm{ \hat{V}^\top \hat{V} \pm \wh{V}^\top V - V^\top V } \\
        &\leq T^{-1} \norm{ \hat{V}^\top \hat{V} - \wh{V}^\top V } + T^{-1} \norm{\wh{V}^\top V - V^\top V } \\
        &\leq T^{-1} \norm{ \hat{V} } \norm{ \hat{V} -  V } + T^{-1} \norm{\wh{V} - V} \norm{ V } \\
        &= O_p\left(\sqrt{\frac{ \log^2(Tp)}{T N^2} } \right).
    \end{aligned}
\end{equation*}
where the first equality follows by adding and subtracting the term; the first and second inequalities follow by the triangle and Cauchy-Schwartz inequalities, respectively; and, the probabilistic bound follows given $\norm{\wh{V}}\lesssim_P \sqrt{kT}$ by the normalization, $\norm{\wh{V} - V^0H }=O_p\left( \frac{\log(Tp)}{N} \right)$ by an analogous argument to Lemma A4; and, $\norm{V} \leq \norm{V^0} \norm{H}=O_p(\sqrt{T})O_p(1)$ by Assumption \ref{assump_latent_fm}(i).
\end{proof}

\begin{lemma}
    Under the assumptions of Lemma A8 and Assumption \ref{assump_inf}(ii), we have $$\norm{T^{-1/2} \wh{V}^\top G - T^{-1/2} V^\top G} = O_p \left( \frac{\log(Tp)}{ N} \right).$$
\end{lemma}

\begin{proof}[Proof of Lemma A12]
\begin{equation*} 
    \begin{aligned}
        \norm{T^{-1/2} \wh{V}^\top G - T^{-1/2} V^\top G} &= T^{-1/2} \norm{ \left(\wh{V} - V \right)^\top \left(V^0 H H^{-1} \eta_0 + \epsilon^g \right) } \\
        &\leq T^{-1/2} \norm{ \wh{V} - V} \norm{V^0} \norm{ \eta_0}  + T^{-1/2} \norm{ \wh{V} - V } \norm{ \epsilon^g  } \\
        &\lesssim_P O_p \left( \frac{\log(Tp)}{ N} \right).
    \end{aligned}
\end{equation*}
where the equality follows from definition of the observable factor model; the inequality follows from the use of the triangle and Cauchy-Schwartz inequalities; and, the probabilistic bound follows as in Lemma A11 with $\norm{\epsilon^g} \lesssim_P \sqrt{T}$ by Assumption \ref{assump_inf}(ii).
\end{proof}

\begin{lemma}
    Under the assumptions of Lemma A12, we have $$ \sqrt{T} (\wh{\eta} - \eta) = O_p \left( 1 \right)$$
\end{lemma}

\begin{proof}[Proof of Lemma A13]
\begin{equation*} 
    \begin{aligned}
        \sqrt{T} (\wh{\eta} - \eta) &= \sqrt{T} \left(\wh{\eta} \pm \wt{\eta} - \eta \right) \\
        &= \left(\frac{\wh{V}^\top \wh{V}}{T} \right)^{-1} \sqrt{T} \frac{\wh{V}^\top G}{T} \pm \left(\frac{V^\top V}{T} \right)^{-1} \sqrt{T} \frac{V^\top G}{T} - \sqrt{T} H^{-1} \eta_0 \\
        &= \left(\frac{\wh{V}^\top \wh{V}}{T} \right)^{-1} \sqrt{T} \frac{\wh{V}^\top G}{T} - \left(\frac{V^\top V}{T} \right)^{-1} \sqrt{T} \frac{V^\top G}{T} \\
        & \qquad + (H^\top)^{-1} \left( \frac{V^{0 \top} V^0 }{T} \right)^{-1} \sqrt{T} \frac{V^{0 \top} \e^g}{T} \\
        &= (H^\top)^{-1} \left( \frac{V^{0 \top} V^0 }{T} \right)^{-1} \sqrt{T} \frac{V^{0 \top} \e^g}{T} + o_p(1) = O_p(1).
    \end{aligned}
\end{equation*}
where the first equality follows from adding and subtracting the infeasible estimator; the second inequality follows from the definitions; the third equality follows from the definition of the model for $G$ and rearranging; the penultimate probabilistic bound follows given the difference between the first two terms is $o_p(1)$ by the results of Lemmas A11 and A12 using an analogous argument as Lemma A10; and, the final probabilistic bound holds given $\norm{H} = O_p(1)$ for invertible matrix $H$, the factors have positive definite second moment matrix given Assumption \ref{assump_latent_fm}(ii), the time series mean of the mean zero random variable $v_{t+1}^0 \e^g_{t+1}$ is $\sqrt{T}$ by CLT Assumption \ref{assump_inf}(iv).
\end{proof}

Define $\Pi_t \coloneqq \sum_{j=1}^s \sum_{j' = 1}^s  \G_{\b,j'} \mathcal{Z}_{t,j,j'} \G_{\b,j}^{\top}$ containing nonstochastic scalar $\mathcal{Z}_{t,j,j'}$ from Assumption \ref{assump_inf}(iv) and the asymptotic variance of the Assumption \ref{assump_inf}(v) joint CLT as
\begin{equation*}
    \begin{aligned}
        \Phi_{11} = \lim_{p,T,N \rightarrow \infty} \frac{1}{T} \E \left [  V^{\top} \e^g \e^{g \top} V \right] &\qquad  
        \Phi_{22} = \lim_{p,T,N \rightarrow \infty} \frac{1}{T} \sum_{t=1}^T \sum_{t'=1}^T \E \left [  \Pi_t v_{t+1} v^{\top}_{t'+1} \Pi_{t'}^\top  \right] \\
        \Phi_{12} &= \lim_{p,T,N \rightarrow \infty} \frac{1}{T} \sum_{t=1}^T \sum_{t'=1}^T \E \left [  v_{t+1} \e^g_{t+1} v^{ \top}_{t'+1} \Pi_{t'}^\top  \right]
    \end{aligned}
\end{equation*}
where asymptotic covariance matrix $\Phi$ is defined as in the logical way.

\begin{proof}[Proof of Theorem 1]
\begin{equation*} 
    \begin{aligned}
        \sqrt{T} \left( \wh{\g}_g - \g_g \right) &= \sqrt{T} \left(\wh{\eta}^\top \wh{\g} - \eta^\top \g \right) \\
        &= (\wh{\eta} - \eta)^\top \sqrt{T} \left( \wh{\g} - \g\right) + \sqrt{T} \eta^\top \left( \wh{\g} - \g\right) + \sqrt{T} \g^\top \left(\wh{\eta} - \eta \right) \\
        &= \sqrt{T} \eta^\top \left( \wh{\g} - \g\right) + \sqrt{T} \g^\top  \left(\wh{\eta} - \eta\right) + o_p(1) \\
        &= \g^{0 \top} \left( \frac{V^{0 \top} V^0 }{T} \right)^{-1}  \frac{\sqrt{T}}{T} \sum_t v^0_{t+1} \e^g_{t+1} \\ 
        & \qquad + \eta^{0 \top} \left(  \frac{\G_\b^{0\top} \bar{Z}^\top \bar{Z} \G_\b^0}{N} \right)^{-1} \frac{ \G_\b^{0\top} \bar{Z}^\top}{N}  \frac{\sqrt{T}}{T} \sum_t Z_t \G_\b^0 v_{t+1}^0   \\
        & \qquad + o_p(1) \\
        &= \g^{0 \top} \left( \frac{V^{0 \top} V^0 }{T} \right)^{-1}  \frac{\sqrt{T}}{T} \sum_t v^0_{t+1} \e^g_{t+1} + o_p(1) \\ 
        & + \eta^{0 \top} \left(  \frac{\G_\b^{0\top} \bar{Z}^\top \bar{Z} \G_\b^0}{N} \right)^{-1} \frac{\sqrt{T}}{T} \sum_t \underbrace{\left( \sum_{j=1}^s \sum_{j' = 1}^s  \G_{\b,j'}^0 \frac{1}{N} \sum_i \E [\bar{z}_{i,j'}] z_{i,t,j} \G_{\b,j}^\top \right)}_{\Pi_t} v_{t+1}^0  \\
        & \rightarrow_d N \left( 0, \s_g^2 \right)
    \end{aligned}
\end{equation*}
where the first equality holds by definition of the estimator and target parameter; the second equality holds by adding and subtracting terms; the third equality holds by Lemma A10 $\sqrt{T} \left(\wh{\g} - \g \right) = O_p(1)$ and Lemma A13 $\wh{\eta} - \eta = O_P (T^{-1/2}) = o_p(1)$; the last equality holds by Lemmas A10 and A13, which leaves the two non-asymptotically negligible terms at rate $\sqrt{T}$ scaled by the associated true parameters $\eta^0$ and $\g^0$; and, the convergence in distribution holds given by the joint CLT assumption \ref{assump_inf}(v) applying the delta method.

Define the following two invertible matrices matrices: $A \coloneqq \lim_{T \rightarrow \infty} T^{-1} \E \left[ V^\top V \right]$ and $B \coloneqq \lim_{p,T,N \rightarrow \infty} \frac{1}{N} \E \left[ \G_\b^\top \Bar{Z}^\top \Bar{Z} \G_\b \right]$.

The asymptotic variance $\sigma_g^2$ is thus given by the delta method:
\begin{equation*}
    \begin{aligned}
    \s_g^2 &\coloneqq \g^\top A^{-1} \Phi_{11} (A^\top)^{-1} \g + \eta^\top B^{-1} \Phi_{22} (B^\top)^{-1} \eta + \g^\top A^{-1} \Phi_{12} (B^\top)^{-1} \eta + \eta^\top B^{-1} \Phi_{12}^\top (A^\top)^{-1} \g.
    \end{aligned}
\end{equation*}
\end{proof}

\newpage
\section{Tables and Figures}\label{s:tables_and_figures}

\begin{table}[h]
\caption{Crypto Asset Characteristics: Descriptive Statistics.}
\includegraphics[width=\textwidth]{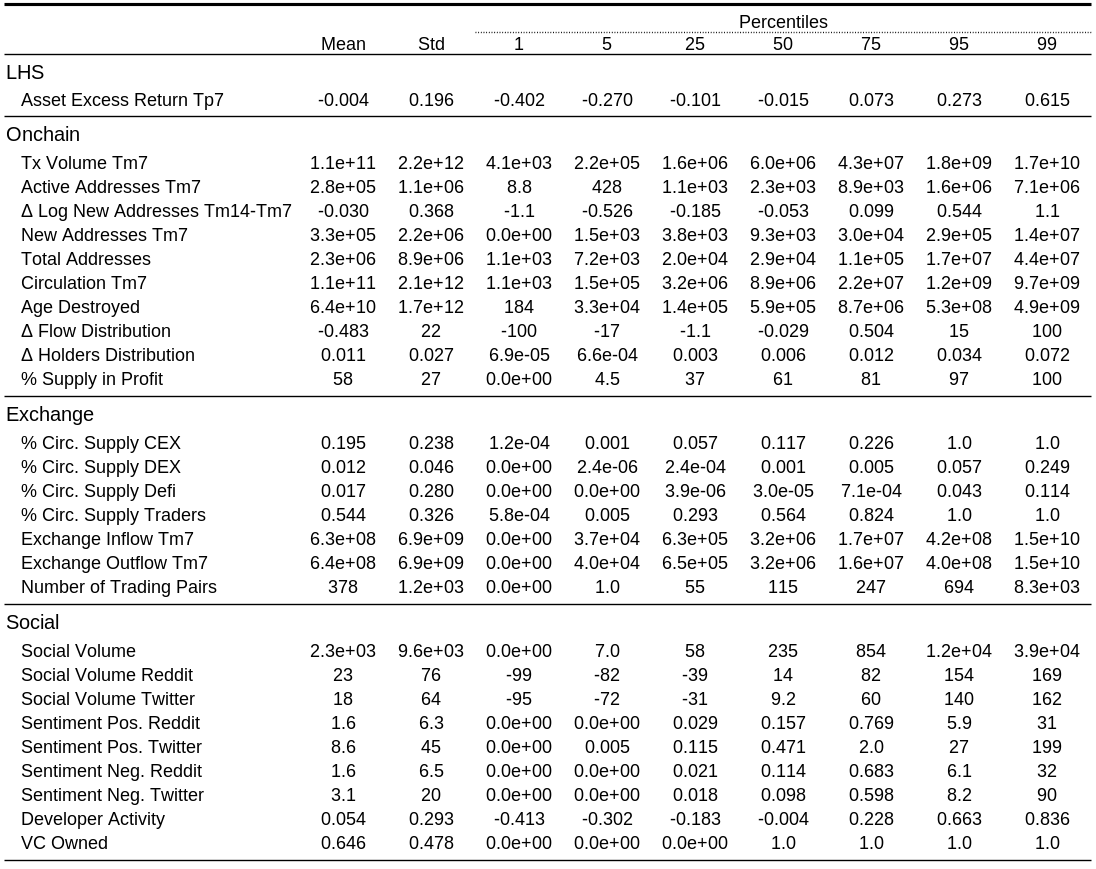}
\note{This table reports the summary statistics from the weekly asset panel for the dependent variable, asset excess returns seven days ahead, and the asset characteristics. For each variable, we report the panel mean, median, standard deviation, and selected percentiles. There are 22,678 asset-weeks from January 7, 2018 to December 15, 2022.}
\label{f:char_desc_stat_panela}\end{table}

\begin{table}[p]
\caption{Crypto Asset Characteristics: Descriptive Statistics (Continued).}
\includegraphics[width=\textwidth]{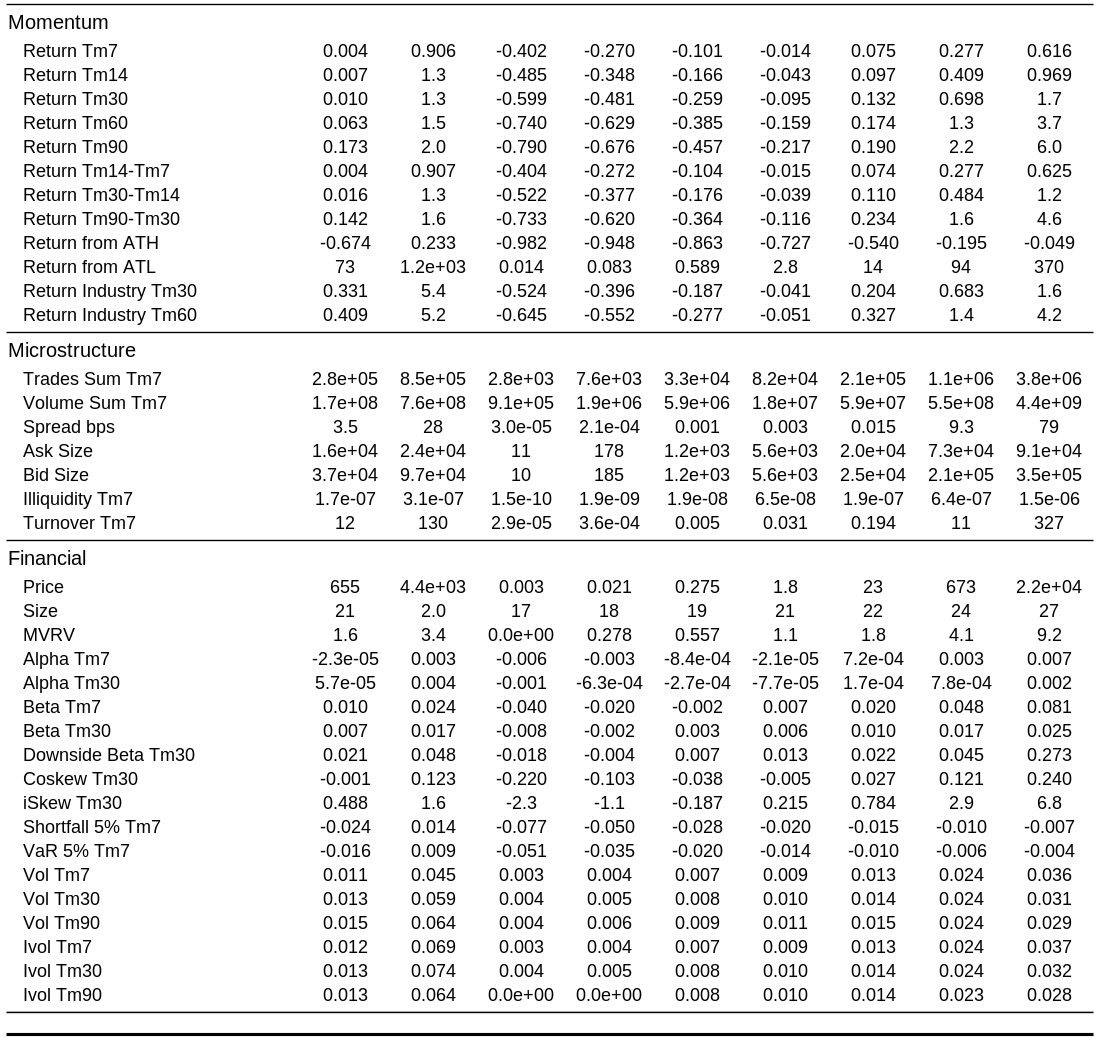}
\note{This table reports the summary statistics from the weekly asset panel for the dependent variable, asset excess returns seven days ahead, and the asset characteristics. For each variable, we report the panel mean, median, standard deviation, and selected percentiles. There are 22,678 asset-weeks from January 7, 2018 to December 15, 2022.}
\label{f:char_desc_stat_panelb}\end{table}

\begin{table}[p]
\caption{Monte Carlo Simulations.}
\includegraphics[width=5.6in]{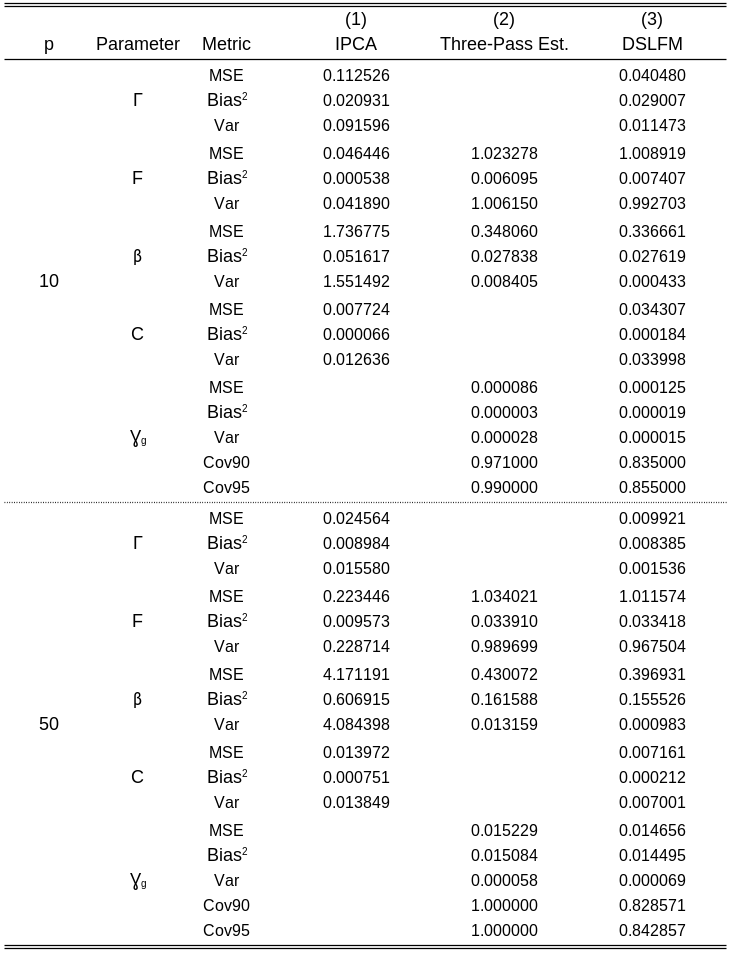}
\note{This table reports Monte Carlo simulations, $S=200,$ for IPCA, Three-Pass Estimators of \cite{giglio2021asset}, and the DSLFM---columns 1, 2, and 3, respectively---for target parameters: latent loadings $\G_\b$, latent-factors $F,$ average factor loadings $\bar{\b},$ latent matrix $C,$ and observable factor risk premium $\g_g.$ The true data-generating process has three factors, $N=500,$ $T=100,$ $p\in \{10,50\},$ and $s=p/10.$ The following metrics are reported: mean-squared error (MSE), bias squared (Bias$^2$), variance (Var), and 90\% and 95\% coverage probabilities (Cov90 and Cov95).}
\label{f:sim}\end{table}

\begin{landscape}
\begin{table}[p]
\caption{Low Dimensional Factor Model Out-of-Sample Returns: Multivariate, PCA, \& IPCA.}
\includegraphics[width=8.6in]{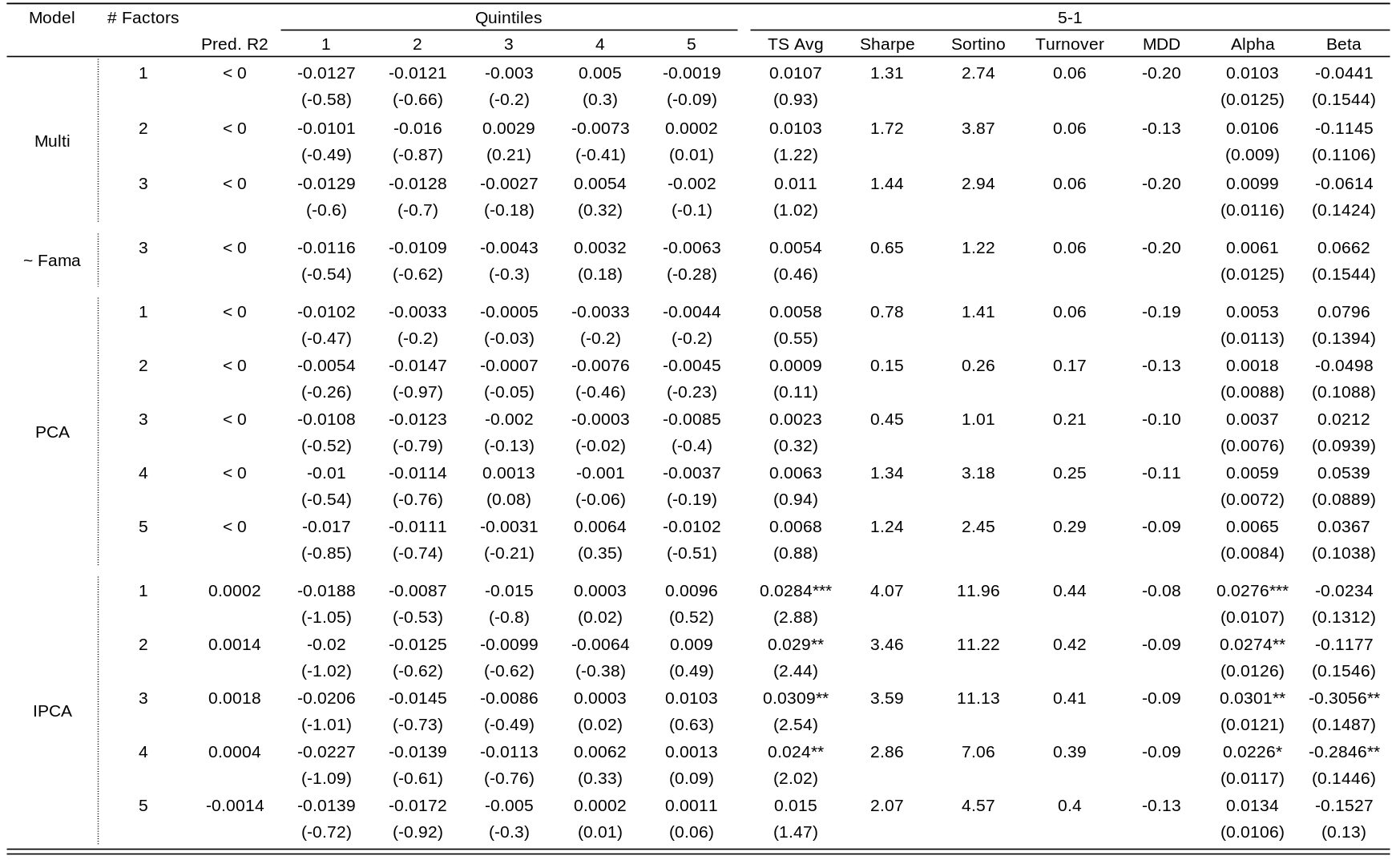}
\begin{minipage}{8.8in} {\scriptsize
This table reports---for multivariate factor models, PCA, and IPCA---the predictive $R^2$, the mean quintile portfolio returns, and portfolio statistics for the 5-1 strategy for July-December 2022, inclusive. For each quintile, the mean returns are the time-series averages of weekly value-weighted portfolio excess returns sorted on each model's predicted returns. 5-1 is the long-short top minus bottom quintile zero-investment portfolio from each model; for which, we report the time-series average weekly value-weighted excess return, annualized Sharpe Ratio, annualized Sortino, weekly turnover, maximum drawdown, and alpha and beta to the CMKT return. \textit{t}-stats are reported below each strategy's point estimates where *, **, and *** denote significance at the 10\%, 5\%, and 1\% levels. Standard errors are Newey-West adjusted using Bartlett's formula for the number of lags. For the multivariate factor model with 1, 2, and 3 factors, the selected characteristics are, respectively: size; illiquidity and size; and, size, 30 day momentum, adn 90 day volatility.
}
\end{minipage}
\label{f:low_dim}\end{table}
\end{landscape}

\begin{landscape}

\begin{table}[p]
\caption{DSLFM Out-of-Sample Portfolio Statistics.}
\includegraphics[width=8.6in]{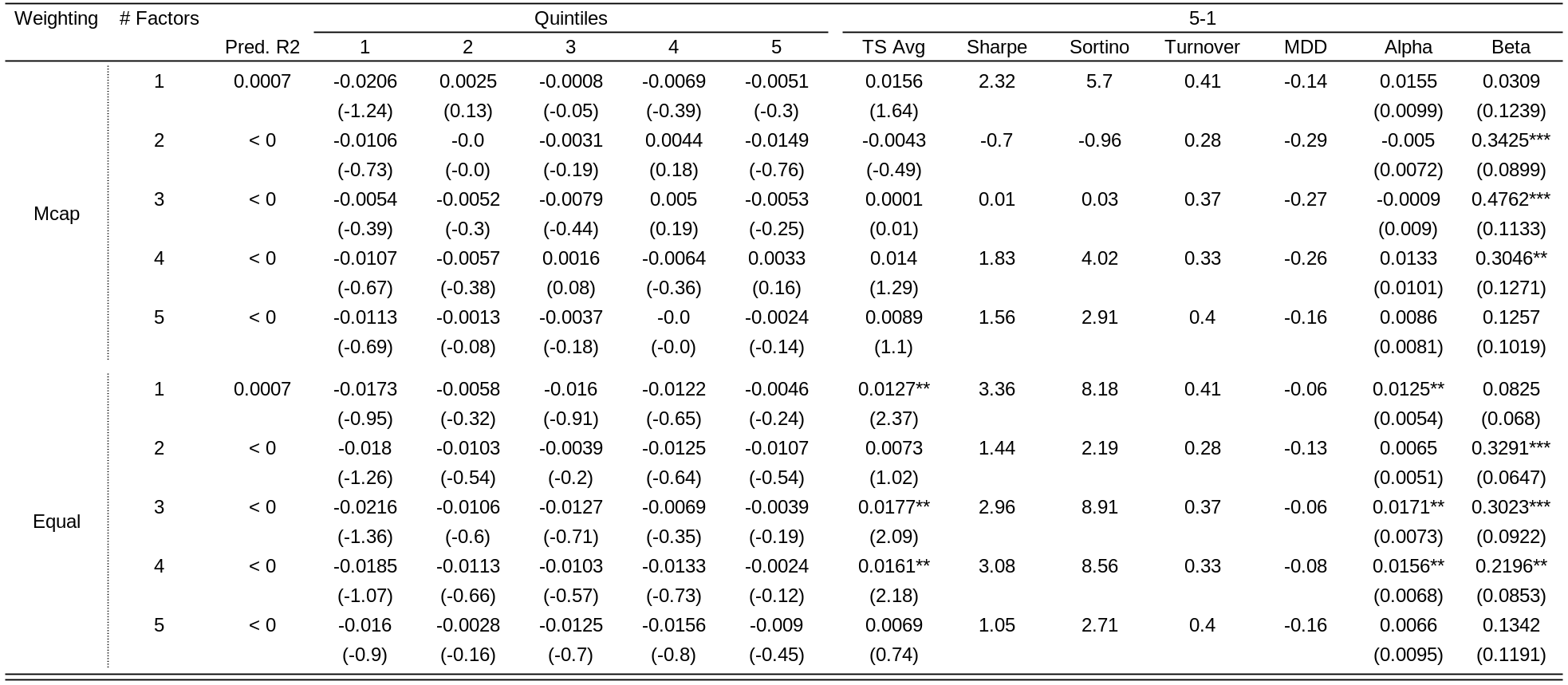}
\begin{minipage}{8.8in} {\scriptsize
This table reports---for the DSLFM with market cap and equal-weighted portfolios---the predictive $R^2$, the mean quintile portfolio returns, and portfolio statistics for the 5-1 strategy for July-December 2022, inclusive. For each quintile, the mean returns are the time-series averages of weekly value-weighted portfolio excess returns sorted on each model's predicted returns. 5-1 is the long-short top minus bottom quintile zero-investment portfolio for each model; for which, we report the time-series average weekly value-weighted excess return, annualized Sharpe Ratio, annualized Sortino, weekly turnover, maximum drawdown, and alpha and beta to the CMKT return. \textit{t}-stats are reported below each strategy's point estimates where *, **, and *** denote significance at the 10\%, 5\%, and 1\% levels. Standard errors are Newey-West adjusted using Bartlett's formula for the number of lags.
}
\end{minipage}
\label{f:dslfm_oos}\end{table}

\end{landscape}

\begin{table}[p]
\caption{DSLFM: Asset Characteristic Significance.}
\includegraphics[width=3.8in]{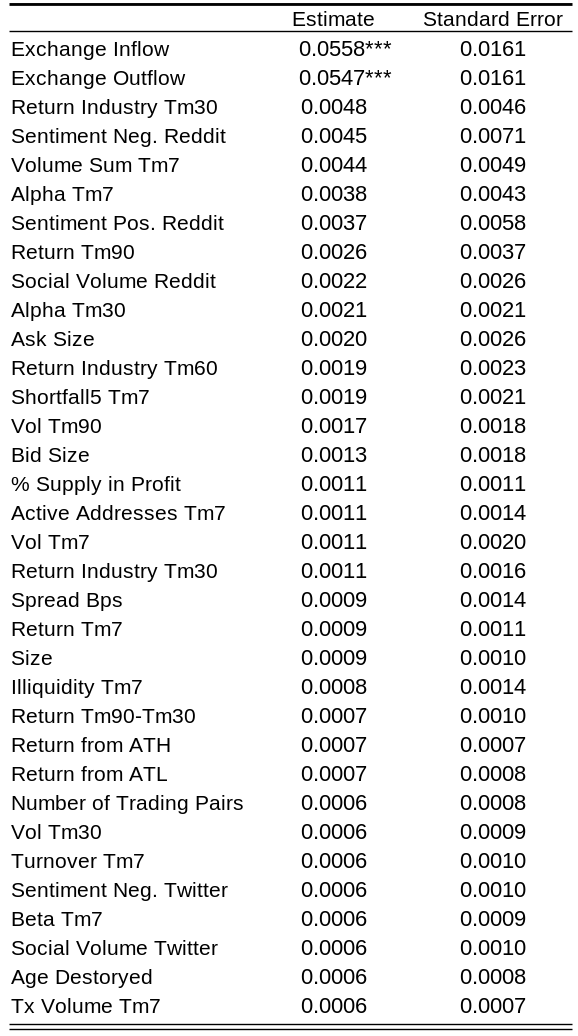}
\note{This table reports estimates of the importance of each asset characteristic to the fitted DSLFM using the test statistic $W_{\G,j} = \G_{\b,j}^\top \G_{\b,j}$ discussed in the theory section. The DSLFM latent loading estimate, i.e. $\wh{\G}_\b$, comes from fitting the DSLFM to the entire weekly panel with hyperparameters selected by the DSLFM CV procedure for the best $k,$ i.e. highest mcap-weighted Sharpe. Standard errors are formed from the standard deviation of the simulated distribution of $\wh{W}_{\G,j}$ using 200 bootstrap draws, for each $j.$ Significance is denoted with *, **, and *** for the 10\%, 5\%, and 1\% levels, respectively. Only characteristics within two orders of magnitude of the maximum estimate are shown, i.e. 34 of the 63 characteristics.}
\label{f:dslfm_char_imp}\end{table}

\end{document}